\documentclass[letterpaper,twocolumn,10pt]{article}
\usepackage{url}
\usepackage{hyperref}
\usepackage{wrapfig}
\usepackage{algorithmic}
\usepackage{graphicx}
\usepackage{textcomp}
\usepackage{tikz}
\usepackage{pgfplots}
\usepackage{multicol}
\usepackage{enumitem}
\usepackage{cuted} 
\pgfplotsset{compat=1.18}
\usetikzlibrary{positioning}
\usepackage[most]{tcolorbox}
\usepackage{mdframed}
\usepackage{pifont}
\usepackage{soul}
\usepackage[table]{xcolor}
\usepackage{longtable,booktabs,array,graphicx}
\definecolor{full}{HTML}{2166AC}    
\definecolor{partial}{HTML}{67A9CF} 
\definecolor{notmet}{HTML}{FDDBC7}  
\definecolor{na}{gray}{0.90}        

\usepackage{caption}
\usepackage{pdflscape} 

\usepackage[nointegrals]{wasysym}   
\usepackage[T1]{fontenc}
\usepackage[utf8]{inputenc}
\DeclareUnicodeCharacter{25CF}{\CIRCLE}       
\DeclareUnicodeCharacter{25D1}{\RIGHTcircle}  
\DeclareUnicodeCharacter{25CB}{\Circle}       
\usepackage{graphicx}  
\newcommand{\FM}{\scalebox{0.9}{\CIRCLE}}      
\newcommand{\PM}{\scalebox{0.9}{\LEFTcircle}}  
\newcommand{\NM}{\scalebox{0.9}{\Circle}}      

\newcommand{\HS}{\scalebox{0.9}{\CIRCLE}}     
\newcommand{\MS}{\scalebox{0.9}{\LEFTcircle}} 
\newcommand{\LS}{\scalebox{0.9}{\Circle}}     
\newcolumntype{Z}{>{\centering\arraybackslash}p{0.65cm}} 
\newcolumntype{T}{>{\centering\arraybackslash}p{0.45cm}} 

\newcommand{\Dfull}[1]{\FM\kern0.2em\cite{#1}} 
\newcommand{\Dpart}[1]{\PM\kern0.2em\cite{#1}} 

\usepackage{pgfgantt}

\newcommand{\fabiha}[1]{\textcolor{orange}{Fabiha: #1}}
\newcommand{\subhav}[1]{\textcolor{orange}{Subhav: #1}}

\usepackage{booktabs}
\usepackage{pdflscape}
\usepackage{adjustbox}
\usepackage{rotating}
\usepackage{tcolorbox}
\usepackage{tabularx}
\usepackage{multirow}
\usepackage{graphicx}
\usepackage{pgfgantt}
\usepackage{xcolor}
\usepackage{svg} 

\usepackage{url}

\usepackage{pgfplots}
\usepackage{longtable}
\usepackage{booktabs}
\usepackage{acronym}
\usepackage{textcase}
\usepackage{xspace}
\usepackage{graphicx}
\usepackage{pgfplots}
\pgfplotsset{compat=1.17}
\usepackage{tikz}
\usepackage{ulem}
\usetikzlibrary{positioning, shapes.multipart, fit, backgrounds}
\def\BibTeX{{\rm B\kern-.05em{\sc i\kern-.025em b}\kern-.08em
    T\kern-.1667em\lower.7ex\hbox{E}\kern-.125emX}}

\usepackage{graphicx,xcolor,caption}
\colorlet{StageGreen}{green!45!white}    
\colorlet{AffectLightRed}{red!25!white}  
\colorlet{ToolRed}{red!80!black}         

\definecolor{DetectBlue}{RGB}{0,102,204}

\usepackage{usenix}

\usepackage{tikz}
\usepackage{amsmath}

\usepackage{filecontents}

\begin{document}

\date{}

\title{Can SOLARBURST go silicon? A systematic analysis on software supply chain
attacks for VLSI design flows}


\author{
{\rm Fabiha Hashmat}\\
Purdue University
\and
{\rm Zeren Li}\\
Purdue University
\and
{\rm Santiago Torres-Arias}\\
Purdue University
}

\maketitle

\maketitle

\begin{abstract}

An attacker targeting the software supply chain (SWSC) of VLSI tools can fully subvert the VLSI design pipeline through purely software level compromises. However, current literature on both SWSC and VLSI hardware security has not explored how software can be used to subvert the security of the VLSI design flow. 
In order to bridge this gap, this paper presents a systematic analysis of the vulnerable surface of VLSI from a SWSC standpoint. We carry this analysis through three major activities. First, we did an empirical analysis of VLSI software suites, focusing on their software composition, security features, and tool interactions within the VLSI design flow. Second, we develop a VLSI-SWSC attack taxonomy model that links software supply chain attacks and their mechanisms to the VLSI attacks they enable, the design-flow stages they affect, and their resulting security outcomes, allowing us to identify 717 possible new attacks, including 60 high feasibility attacks supported by evidence collected in our empirical analysis. Third, we identify what possible security outcomes can be achieved by an attacker from these attacks and produce a comprehensive VLSI SWSC threat model. We evaluate the feasibility of the attacks identified by our taxonomy across VLSI design stages and assess existing defenses to determine which attacks are mitigated and which remain unaddressed. In addition, we present case studies using a reference VLSI design flow consisting of open-source VLSI tools and highlight attacks identified through our taxonomy on the VLSI SWSC uniquely from a software perspective.
\end{abstract}

\section{Introduction}

The software supply chain (SWSC) of the VLSI supply chain is a combined network of software artifacts (e.g., design tools, source code, and open-source or third party dependencies), and software driven processes (e.g., build pipelines, deployment configuration/environment) involved in the design, transformation, optimization and validation of VLSI designs~\cite{efablessMPW}.
Modern VLSI design projects (e.g., Google--SkyWater Open MPW program~\cite{skywater2020,googleSilicon, efablessMPW}, Riscduino \cite{riscduino2021, chipIgnite}, and Microwatt 64-bit OpenPOWER core \cite{microwattGithub}) increasingly rely on a SWSC composed of open-source tools spanning multiple stages, these include Yosys, OpenROAD, Magic and the OpenLane. This reliance is consistent with industry trends: over 90\% of codebases in the computer hardware and semiconductor domain include open-source components, which constitute a significant fraction of the overall software stack ~\cite{ossra2023}.
Similarly, commercial VLSI design projects such as Cadence also relies on the use of open source software or IP licensed from third parties in their products \cite{cadence10k2024}.




While these projects benefit from open-source and third-party software for accelerated development, they also introduce a wholly new attack surface into the VLSI design process. 
Large-scale industry studies show that open-source components are pervasive and frequently vulnerable: the 2024 Open Source Security and Risk Analysis (OSSRA) report by Synopsys found that 84\% of audited codebases contained at least one known vulnerability \cite{synopsys2024ossra}. In parallel, SWSC attacks targeting open-source dependencies have increased sharply in recent years~\cite{sonatype2022}. 

Further, according to Cadence 2024 Annual Report \cite{cadence10k2024}, they admitted the use of open source solutions in their products may cause vulnerabilities that are in turn exploited by attackers. 
When those open source and third-party components with known vulnerabilities are integrated into VLSI toolchains as trusted software artifacts, design-level flaws may be introduced, and their misbehavior can propagate across the entire VLSI design pipeline. 
An attacker targeting the SWSC of these VLSI software suites maybe able to fully subvert a VLSI pipeline uniquely from a software perspective. 
This highlights a fundamental challenge in securing VLSI design process SWSC -- how can we ensure the trustworthiness of a VLSI design when misbehaving open-source software within its toolchain can compromise the entire pipeline?

However, literature on both SWSC \cite{kande2022thehuzz,dessouky2019hardfails} and VLSI security \cite{miftah2024rtl,srivastava2024scar,datta2022security,meng2021rtl} has not explored the ways in which software can be used to subvert the security of the VLSI supply chain. 
For example, methodologies like split manufacturing \cite{gohil2021games}, which are effective in protecting against hardware Trojans, IP theft \cite{chakraborty2020hardware,han2023functeller}, and unauthorized reproductions \cite{vedula2018system}, prompt the inquiry — can these threats also be conducted through a software only vector? 
This same vector can also enable IP theft and unauthorized reproduction, since VLSI software suites are tightly interconnected and depend on external components such as third-party IP cores.
In practice, this occurs through software-level compromise of tool binaries, plugins, or shared libraries that handle intermediate design artifacts, which often contain proprietary IP.

To address this research gap, we aim to answer: how an attacker can subvert a VLSI design process, uniquely from a software perspective, and what can they achieve by doing so? 
The main contributions of the research are:\\
\textbf{Systematic study of reference VLSI software composition:} we analyze select VLSI software suites to identify their SWSC, with a particular focus on open source components and libraries utilized during ``pre-silicon'' phases: VLSI design, testing, and synthesis.\\
\textbf{VLSI-SWSC attack taxonomy:} we develop a VLSI-SWSC attack taxonomy that links SWSC attacks and their mechanisms to the VLSI attacks they enable, the design-flow stages they affect, and their resulting security outcomes, allowing us to identify 717 possible new attacks, including 60 high-feasibility attacks supported by evidence collected in our empirical analysis.\\
\textbf{Exhaustive study of possible security outcomes:} with the two above, we identify what possible outcomes can be carried out from these attacks and use them to construct a comprehensive VLSI-SWSC threat model.\\
\textbf{Defense Analysis:} Leveraging the CIS SWSC Security Guide, we evaluate the coverage of existing defenses against VLSI SWSC attack-path classes and identify gaps in current mitigation strategies.\\
\textbf{Case Study:} We evaluate our threat model on a representative VLSI design flow consisting of open-source tools by implementing software-only attacks selected from our VLSI-SWSC attack taxonomy, focusing on vectors introduced through third-party libraries and dependencies. We discuss one speculative attack scenario derived from the same taxonomy.

Our analysis reveals that software dependencies embedded within VLSI software suites can serve as direct attack vectors capable of subverting the integrity of the entire VLSI design process. Through our empirical study and VLSI-SWSC attack taxonomy, we identify 717 possible new attacks, including 60 high-feasibility attacks supported by evidence collected in our empirical analysis, and characterize how these attacks can affect different VLSI design stages. Our results show that while certain classes of attacks are partially mitigated by existing industry practices and current literature, significant coverage gaps remain particularly for attacks emerging from unvalidated open-source components and inter-tool interactions. By mapping these model identified attacks to defense controls outlined in the CIS SWSC Security Guide, our study reveals critical blind spots in existing VLSI SWSC protections and highlights where future defenses must improve. 
\begin{figure}[htbp]
  \centering
\includegraphics[width=0.85\linewidth, trim=80 0 67 0]
{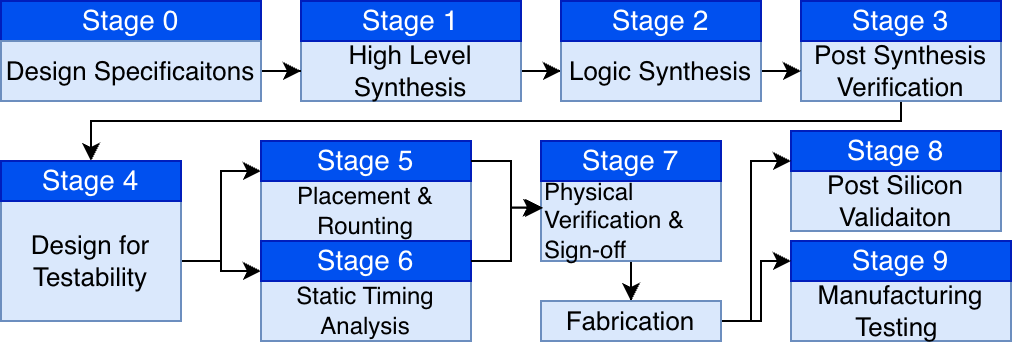}
  \caption{VLSI Design Process}
  \label{step}
\end{figure}

\section{Background and Research Gap}
\label{Background}


This section introduces the terminology used throughout the paper, outlines the software supply chain of the VLSI design process, characterizes SWSC attack phases, and identifies the research gap that motivates our study.

\subsection{Terminology}
Multiple organizations define SWSC terminology, including the National Institute of Standards and Technology (National Institue of Standards and Technology, NIST) through SP 800-161 Rev. 1 and the Secure Software Development Framework (SSDF), the Open Source Security Foundation  (OSSF) threat classifications, the Cybersecurity and Infrastructure Security Agency (CISA), the European Union Agency for Cybersecurity (ENISA) ICT supply chain risk models, the Center for Internet Security (CIS) SWSC Security Guide. While these differ in scope and intended audience, they consistently describe the SWSC in terms of software artifacts, build and integration process. Among these NIST consolidates all of these core concepts within a single, unified framework, providing sufficient coverage of the terminology required for our analysis. Prior software SWSC research commonly adopts NIST terminology when formalizing attack surfaces. We therefore follow NIST terminology as a single reference point to remain consistent with existing literature and to avoid redundant comparison of largely overlapping definitions.

Following are the definitions of basic supply chain terms being used henceforth.\\
\textbf{Software Supply Chain:} NIST SP 800-204D ~\cite{nist204d} defines a software supply chain as “a collection of steps that create, transform, and assess the quality and policy conformance of software artifacts. These steps are often carried out by different actors who use and consume artifacts to produce new artifacts”\\
\textbf{Supply Chain Attack:}
NIST SP~800-161~Rev.~1 ~\cite{nist800161} defines this as
``an occurrence within the supply chain whereby the confidentiality,
integrity, or availability of a system or the information the system
processes, stores, or transmits is jeopardized.''\\
\textbf{SWSC Attack:}
NIST SP~800-204D ~\cite{nist204d} defines this as
``a malicious party tampering with the steps, artifacts, or actors
within the chain to compromise downstream consumers of a software
artifact.'' Such attacks typically follow a three-stage process \cite{ishgair2024sok}:
(i) artifact, step, or actor compromise,
(ii) propagation through the chain, and
(iii) exploitation to achieve attacker goals (e.g., data exfiltration,
cryptojacking).\\
In this context, \textbf{actors} represents the entities that perform activities in the supply chain (e.g., developers, or build systems); \textbf{steps} refer to the software-development or delivery processes (e.g., build, test, package, deploy); and \textbf{artifacts} are the software objects exchanged between these steps (e.g, source code, binaries, packages, or container images). \\
\textbf{CIA Triad:}
As we know, confidentiality, integrity, and availability (CIA) form the core security objectives; we adopt the NIST definitions from NIST SP~800-12~Rev.~1~\cite{nist80012r1}.
Confidentiality—preserving authorized restrictions on access and disclosure, including protection of privacy and proprietary data;
Integrity—guarding against improper modification or destruction and ensuring authenticity; and
Availability—ensuring timely and reliable access.\\




\subsection{SWSC of VLSI Design Process}

Following NIST's definition of a software supply chain, we consider the VLSI SWSC as the software-supply-chain steps and artifacts involved in the VLSI design flow, where EDA software consumes, transforms, and produces design artifacts across successive design stages. This scope includes the EDA tools and their software components and dependencies used from design specification through manufacturing testing.

Various tasks are involved in creating an electronic device, including laying out functional units of a device (i.e., logical portions of the device), synthesizing that design into a low-level gate design, and ensuring the quality of the end-result  (e.g., by ensuring electrical/timing properties are held), among others.
\cite{vai2000vlsi} \cite{52217, bhunia2018hardware}
Below, we elaborate on the most common activities performed on the VLSI design flow. 
Figure~\ref{step} provides a graphical overview of the VLSI design process steps, compiled through a study of tool vendors such as Cadence~\cite{cadence_vlsi_fundamentals}) as well as academic sources like~\cite{basu2019cad}. 
In the context of this paper, we focus on the pre-silicon phase-from design specification to physical verification and sign off.
We list the tools within this flow in Table ~\ref{tab:tool_system_capabilities}.



\noindent We examine pre-silicon design workflows because the software used at this stage—synthesis, simulation, verification, and timing analysis—pulls in a diverse set of third-party components (e.g., SAT/SMT solvers, parsing libraries, numerical optimization engines, scripting runtimes) \cite{qiu2025aieda,anysilicon_open_source_eda}.\\ 
\textbf{Design Specification:} Defines high-level functional requirements and identifies the set of Intellectual property blocks to be used. 
These IPs are often sourced from third-party vendors (3PIPs)~\cite{bhunia2018hardware}, and pose a security risk if malicious or unverified, potentially introducing backdoors or undocumented behaviors from the outset.\\
\textbf{High-Level Synthesis (HLS):} converts software-like specifications (e.g., C, MATLAB) into hardware level descriptions (e.g., Verilog). A compromised HLS tool or script can misallocate resources, insert convert logic or degrade parallelism-introducing inefficiencies or malicious behavior into the RTL (Register Transfer Level).\\
\textbf{Verification:} confirms functional correctness after HLS using
simulation or formal methods. Formal techniques—such as equivalence checking, model checking, and theorem proving—verify
behavior against specifications. If verification fails to capture transformations, injected logic or altered properties may
remain undetected\\
\textbf{Logic Synthesis:} converts RTL into a gate-level \emph{netlist} using standard cell libraries with delay and power data. 
It includes logic simplification, synthesis, technology mapping and optimization. 
Malicious scripts or tool misuse can modify logic or insert hidden structures, risking functional correctness.\\
\textbf{Post-Synthesis Verification:} ensures the synthesized gate-level netlist matches the RTL design. 
Equivalence checking is used to detect unintended changes. 
Compromising this step may fail to catch logic tampering/synthesis induced deviations.\\
\textbf{Design for Test (DFT):} adds test structures like scan chains and JTAG for post-silicon debugging.
It includes placement and routing (floor-planning, clock tree synthesis, etc.) and static timing analysis to ensure timing correctness.
Malicious manipulation can misreport timing paths, enabling timing-based side-channel attacks, and expose internal states or introduce access points (e.g., a hidden JTAG port).

\subsection{Software Supply Chain Security}

SWSC security have two related but distinct categories: vulnerable supply chain and deliberate supply chain attacks. The first category covers accidental weakness (vulnerable dependencies or poor development practices) that adversaries may later exploit. While second category,deliberate supply chain attacks result in by intervention in trusted tools, dependencies, build processes to alter downstream supply chain steps and artifacts. In this paper, we are focused on the later scenario.

Software Supply Chain Attacks~\cite{torresarias2020intoto, ladisa2023sok, okafor2022sok}, are devastating attack vectors that can fully subvert a software product without exploiting vulnerabilities in the software itself. Such attacks typically involve three phases: 
\textbf{injection}, where an attacker compromises a dependency or tool (e.g., a compiler) to insert malicious code; 
\textbf{propagation}, where the modified artifact spreads through the supply chain to reach the target; and 
\textbf{exploitation}, where the attacker leverages the compromised artifact (e.g., a backdoored binary) to achieve their goals.

As recent incidents have shown, SWSC attacks have become increasingly prevalent, impacting organizations across multiple sectors, including government systems \cite{wiredSolarWinds2021}.


\subsection{Research Gap:}

With this in mind, we posit the following research question:
\textbf{In what ways can an attacker subvert a VLSI design process, uniquely from a SWSC perspective?} \\
\textbf{RQ 1:} What are the key characteristics of VLSI
        software suites and nature of the software vulnerabilities that are often introduced in VLSI software through their supply chain, and how can we systematically categorize their roles, capabilities, and interactions within the design process? \hyperref[sec:software suites]{(Section 4)}\\
\textbf{RQ 2:} What are the potential security outcomes of SWSC attacks, and how can we systematically categorize these outcomes to ensure a comprehensive and exhaustive security taxonomy? \hyperref[sec:ThreatModel]{(Section 6)}\\
\textbf{RQ 3:} What defense mechanisms—analyzed using the CIS SWSC Security Guide—currently exist to mitigate VLSI-SWSC attack model classes? \hyperref[sec:feasibility_defense]{(Section 7)}\\

\section{Methodology}

To answer these research questions, we conduct a structured analysis of the current literature and an in-lab empirical analysis of VLSI software suites using standard software analysis techniques.
Succinctly put, we aim to understand the state of the art in the VLSI design flow to identify the tools used within it, then develop a comprehensive VLSI-SWSC attack taxonomy of software-only entry points for attackers in this chain, linking SWSC attacks and their mechanisms to the VLSI attacks they enable, the design-flow stages they affect, and their resulting security outcomes. To evaluate the feasibility of the attack paths identified by our taxonomy model, we use empirical evidence from our VLSI software analysis to classify the synthesized paths as high, medium, or low feasibility. We then assess how existing defenses align with these attack-path classes.

Multiple software supply chain security frameworks exist, such as NIST SSDF, NIST SP 800-161 Rev.1, SLSA, CISA's SBOM guidance, and the CIS SWSC security guide. Each offers different scopes and levels of abstraction, from high-level organizational governance (e.g, SSDF) to developer and build-system level operational controls (e.g, CIS, SLSA). Without a consistent reference model, evaluating attacks and defenses becomes difficult. Therefore, we select the CIS SWSC Security Guide because, among existing frameworks, it offers the most implementation-oriented control families that map directly to the developer, dependency, and build behavior present in VLSI software suits. Whereas SSDF and SP 800-161 provide high level policy guidance, and SLSA focuses narrowly on buid integrity provenance, CIS provides controls spanning source integrity, dependency validation, build hardening, and artifact verification which precisely covers the surfaces exploited in software-centric VLSI threats. 
We then mapped all attacks to the CIS SWSC Security Guide \cite{cis_software_supply_chain_guide} - a consensus-based framework outlining measurable practices for securing SWSC - to quantify defense coverage across the identified attack-path classes Section~ (\ref{tab:defense_mapping_cis}). 

We further test this threat model through a case study using a representative VLSI design flow composed of open-source software. We implement three attacks selected from our VLSI-SWSC attack taxonomy (P376, P703, and P616) and separately discuss one additional taxonomy-derived attack scenario (P624) as speculative.
For open-source tools we used Yosys (a logic synthesis tool) ~\cite{Yosys}, OpenLane (a toolchain providing RTL-to-GDSII implementation flow)~\cite{openlane2020} and Magic \cite{Magic}. We selected these tools based on their active community development, citation frequency in academic literature, and inclusion in widely used open-source VLSI flows such as OpenLane and the Google–SkyWater Open MPW program \cite{skywater2020}.




\begin{figure}[t]
  \includegraphics[width=1.0 \linewidth]{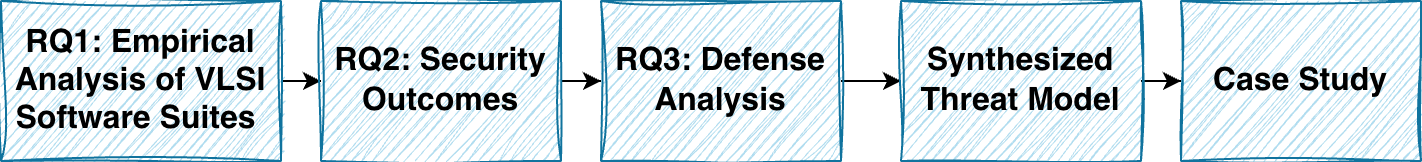}

  \caption{The research methodology. We first explore the vulnerable surface from VLSI software, incorporate an attack feasibility assessment and a defense analysis guided by the CIS framework, and then tie this surface to security-relevant outcomes. We finally synthesize a threat model using this information and test the model against a VLSI design flow.}
  \label{methodology1}
\end{figure}




\textbf{Mapping Software Tools For VLSI Design (Section \ref{sec:software suites}):} In order to concretize the possible operations carried by software within the VLSI design flow, we study popular open source  VLSI software suits which are commonly used in academic and industry workflows.
 
Open source VLSI design process tools are gaining traction as a cost effective, easy access to education and research, and the transparent, open collaborative and community driven nature provides an alternative to proprietary solutions, which often pose barriers to accessibility due to licensing restrictions and vendor lock-in~\cite{voltpost2024oseda}. 
Notably, initiatives like DARPA's Electronics Resurgence Initiative have actively supported open-source EDA development.
Moreover, strategic assessments of semiconductor resilience now recognize that long-term trends such as "open-silicon" -encompassing open source EDA, IP and PDKs- warrant global support, especially in response to geopolitical constraints on proprietary EDA exports~\cite{kleinhans2022opensilicon}. 


Further, we lean on literature from~\cite{basu2019cad} and~\cite{ladisa2023sok} that explore conceptual models for the VLSI design process. This way, we are able to categorize individual tools within each suite, within a particular application domain.
Doing this will enable us to develop three pieces of knowledge:
first, a clear understanding of tool capabilities (i.e., what is expected from a tool, behaviorally). This includes the expected inputs/outputs and transformations a tool performs. Without baseline understanding of behavior, we cannot determine whether an attack/anomaly reflects normal tool functionality.
    
    Second, a concrete notion of data flow between tools (i.e., which tools produce software artifacts used by other tools). This data flow structure determines where supply chain vulnerabilities can propagate.
    
    Third, an overview of which vendors provide which tools, because we assume that companies may use a combination of vendors to fully satisfy their design activities. This context allows us to reason about cross-tool trust relationship and vendor-specific risk surfaces.\\
\textbf{Identifying Security Outcomes: (Section \ref{sec:outcomes}):} 
We carry out a systematic exploration of possible security outcomes and identify security expectations such as data privacy (i.e., protecting intellectual property), data integrity (i.e., preventing hardware trojans), and availability (i.e., preventing sabotage of finalized designs).\\
We use these security outcomes, together with the identified software-side attack vectors, exploitation mechanisms, and VLSI design stages, to construct our VLSI-SWSC attack taxonomy.
In other words, by understanding the security expectations and the capabilities of a tool, we are able to answer research question RQ2 in order to achieve an outcome, what class of tools should an attacker compromise using a SWSC attack?\\
\textbf{Threat Model Synthesis (Section \ref{sec:ThreatModel}):}
Lastly, we employ the information from the two previous activities to synthesize and formalize a threat model by developing a taxonomy-based model that outlines the \emph{degrees of access} that an attacker require to achieve a security outcome.\\
In other words, we combine tool capabilities, with security expectations, and attacker access, to derive all possible SWSC related entry-points for an attacker on each individual target.\\
This taxonomy-based model also allows us to determine strategies to enumerate and explore the vulnerable surface of any particular instantiation of a VLSI design software.\\
\textbf{Attack and Defense Analysis (Section \ref{sec:feasibility_defense}):}
Building on the VLSI-SWSC attack taxonomy, we evaluate the feasibility of the identified attacks and examine the extent to which existing defenses address them. Given its comprehensive coverage of software supply chain controls and industry adoption, we employ the CIS Software Supply Chain Security Guide as a structured framework to analyze countermeasures against software supply chain threats in VLSI flows. This structured analysis highlights critical coverage gaps in mitigating SWSC threats within VLSI design flows, guiding the development of our defense mapping framework. We further evaluate the feasibility of synthesized attack paths using empirical evidence and assess their coverage under existing software supply chain security controls
\textbf{Evaluation (Section \ref{sec:casestudy}):} 
Finally, we utilize the model derived above to select representative attacks for evaluation on an open-source VLSI design flow. Specifically, we implement three attacks selected from our VLSI-SWSC attack taxonomy—P376, P703, and P616, covering confidentiality, availability, and integrity outcomes, respectively, and separately discuss one taxonomy-derived speculative scenario, P624. Using these scenarios, we demonstrate how compromises of VLSI tools and their dependencies can affect design artifacts and downstream stages of the VLSI design flow.

\section{Empirical Analysis of VLSI software suites}
\label{sec:software suites}


This section presents an empirical analysis of VLSI software suites to understand \textit{why} and \textit{how} their roles, capabilities, interactions, and dependencies form a critical SWSC risk surface within the VLSI design process. We address this through the following research questions.

\noindent \textbf{RQ 1.1} What types of software are used in carrying out VLSI design process and how do these tools interact with each other?\\
\textbf{RQ 1.2} What are the capabilities of these tools with regard to the software artifacts being produced or used?\\
\textbf{RQ 1.3} What types of software libraries are utilized by these software tools, and how do these dependencies contribute to their vulnerable risk surface?\\
We address each sub-question in the sections below.
\begin{table*}[h]
\centering
\footnotesize
\caption{Capabilities and System Level Interactions of VLSI Software Suites. File System Access symbols: \ding{108} = Present, \PM = Partial, \NM = None. \ding{51} = Yes, \ding{55} = No.}
\label{tab:tool_system_capabilities}
\renewcommand{\arraystretch}{1.2}
\begin{adjustbox}{max width=\textwidth}
\begin{tabular}{|p{2.5cm}|c |c|c|c|c|c|c|c|}
\hline
\textbf{Design Step} & \textbf{Tool} & \multicolumn{4}{c|}{\textbf{File System Access}} & \textbf{CIS Technique}
 & \textbf{Third-party Tool Invoked} & \textbf{Shell Injection} \\ 
\cline{3-6}
 & & \textbf{Artifact} & \textbf{Log} & \textbf{Config} & \textbf{Other} & & & \\ \hline

High Level Synthesis & LegUp & \ding{108} & \NM & \PM & \NM & Code Risks & \ding{51} & \ding{55} \\ \hline

High Level Synthesis & Bambu & \ding{108} & \NM & \PM & \NM & Code Risks & \ding{51} & \ding{55} \\ \hline

Verification & Symbiyosys & \ding{108} & \ding{108} & \PM & \NM & Third-Party & \ding{51} & \ding{55} \\ \hline

Logic Synthesis & Yosys & \ding{108} & \NM & \ding{108} & \PM & Code Risks & \ding{55} & \ding{55} \\ \hline

Logic Synthesis & ABC & \ding{108} & \NM & \ding{108} & \PM & Pipeline Instructions & \ding{55} & \ding{55} \\ \hline

Placement \& Routing & Magic & \ding{108} & \NM & \ding{108} & \PM & Third-Party & \ding{51} & \ding{51} \\ \hline

Placement \& Routing & Qrouter & \ding{108} & \NM & \PM & \PM & Third-Party & \ding{55} & \ding{51} \\ \hline

Static Timing Anal. & OpenTimer & \ding{108} & \ding{108} & \NM & \NM & Code Risks & \ding{55} & \ding{55} \\ \hline

Design for Testability & Fault & \ding{108} & \ding{108} & \NM & \NM & Access to Artifacts & \ding{55} & \ding{55} \\ \hline

Manufacturing Test & Atalanta & \ding{108} & \PM & \NM & \NM & Code Risks & \ding{55} & \ding{55} \\ \hline

\end{tabular}
\end{adjustbox}
\end{table*}


\paragraph{RQ 1.1: Tools Across VLSI Design Process and Their Interactions}
To answer RQ 1.1, we reviewed vendor documentation, open-source repositories, and academic literature to map the key open-source and proprietary tools used across each stage of the VLSI design process and analyze their interactions. This mapping reveals that key stages—especially logic synthesis and verification—rely heavily on open-source tools. This concentration matters because a compromise introduced through an open-source tool at these stages can persist in the generated design artifacts and propagate to subsequent stages, regardless of whether the downstream flow is implemented using open-source, proprietary, or mixed EDA toolchains.

\begin{tcolorbox}[
  colback=green!5,
  colframe=green!60!black,
  title=RQ 1.1 Key Findings,
  fonttitle=\bfseries,
  boxrule=0.5pt,
  arc=2pt,
  left=4pt,
  right=4pt,
  top=2pt,
  bottom=2pt,
  enhanced,
  sharp corners,
  fontupper=\fontsize{8pt}{9pt}\selectfont
]
Open-source tools support most pre-silicon design stages, with the highest concentration in logic synthesis, verification, and placement \& routing, critical to design integrity and optimization as shown in Table \ref{tab:shared_static_tpl}. The growing contributors base e.g., Yosys with over 552 contributors (contributor counts are derived from public Github statistics) signals a rising shift towards community-driven alternatives in VLSI design.
\end{tcolorbox}

Table~\ref{tab:tool_system_capabilities} shows the tools utilized in VLSI design process. 




\paragraph{RQ 1.2 Capabilities of tools with regard to the software artifacts being produced or used}
To evaluate these capabilities, we employed a multi-pronged methodology combining dynamic syscall tracing, static dependency inspection, and privilege analysis.\\

\begin{tcolorbox}[
  colback=green!5,
  colframe=green!60!black,
  title=RQ 1.2 Key Findings,
  fonttitle=\bfseries,
  boxrule=0.5pt,
  arc=2pt,
  left=4pt,
  right=4pt,
  top=2pt,
  bottom=2pt,
  enhanced,
  sharp corners,
  fontupper=\fontsize{8pt}{9pt}\selectfont
]
Across the tools analyzed in Table \ref{tab:tool_system_capabilities}, 100\% operated at user-level privileges and none initiated network connections. However, 80\% supported external script execution or third-party tool invocation. 
\end{tcolorbox}
\textbf{System call tracing:} We used \texttt{strace} and \texttt{ltrace} to observe key system-level interactions such as file access (\texttt{openat}, \texttt{read}, \texttt{write}), shell invocations (\texttt{execve}, \texttt{system}), and inter-process communication (\texttt{mmap}, \texttt{socket}, \texttt{shmget}).\\
\textbf{Dependency inspection:} Using \texttt{ldd} to capture dynamically linked libraries and detect obsolete or vulnerable \texttt{.so} versions.\\
\textbf{Privilege examination:} Using \texttt{/proc/<pid>/status} and \texttt{pscap} to verify whether elevated capabilities (e.g., \texttt{CAP\_SYS\_ADMIN}) are required.\\
\textbf{Security feature identification:} We examined official documentation, help menus (e.g., \texttt{--help}, \texttt{-h}), and runtime logs to identify any built-in security options such as sandboxing, audit logging, privilege dropping, or strict parsing modes (e.g., \texttt{-e}, \texttt{-s}, \texttt{-x}.)
The tools were exercised with minimal but realistic workflows (e.g., loading verilog files, running synthesis) to simulate standard use cases.\\
Table~\ref{tab:tool_system_capabilities} shows these observations. 
Our findings reveal that all analyzed tools operate at user-level privileges without requiring special system capabilities.
Most tools lack essential security mechanisms such as \texttt{input validation}, and \texttt{privilege confinement}. \textbf{Yosys}, \textbf{SymbiYosys}, and \textbf{LegUp} operate in a \texttt{path-confined} mode without invoking external tools or shell commands, consistent with lower exposure in CIS \texttt{Code Risks} and \texttt{Third-Party} categories. 
\textbf{ABC} and \textbf{Bambu} exhibit similar isolation levels but rely on user-level configuration files or compilers, corresponding to exposure captured under the CIS \texttt{Pipeline Instructions} and \texttt{Code Risk} categories.
\textbf{Magic} and \textbf{Qrouter} accept \texttt{external Tcl scripts} without validation and lack syscall restrictions. This behaviors reflects exposure within the CIS \texttt{Third-Party} category.
While \textbf{Magic} invokes third-party plotting utilities, \textbf{Qrouter} supports \texttt{exec} within scripts, posing clear \texttt{code injection} risks which is captured under the CIS \texttt{Third-Party} category.. 
These tools present moderate exposure due to \texttt{unvalidated scripting} and loose integration boundaries.
\textbf{OpenTimer}, \textbf{Fault}, and \textbf{Atalanta} operate with minimal system interaction, lack external tool invocation, and show no evidence of privileged syscall use. 
Their execution is limited to local I/O with a narrow \texttt{syscall surface}. This behavior corresponds to lower exposure within the CIS \texttt{Code Risks} and \texttt{Access to Artifacts} category.

\paragraph{RQ 1.3 Software Libraries utilized by VLSI Software Suites}
We identified third-party libraries in VLSI software using a combination of static analysis and dynamic inspection techniques. 
This included parsing build scripts, analyzing binary symbol tables using off-the-shelf tools (e.g., \texttt{ldd}) as well as custom tools, to performing source level inspection and trace both shared and statically linked external code. 
Concretely, we set off to identify the following measures:\\
\textbf{Static TPL Library Metadata, Ownership in VLSI Software Suites} 
        To measure the risk surface from a SWSC level, we identified the amount of external code and functions used by these tools.
        We sought to identify code that was statically integrated in these tools first.
        We measured these in low-level instructions (i.e., the percentage of code that exists in the tool's binary).\\
\textbf{Percentage of Third-Party Shared Libraries Among All Shared Libraries Used by Each VLSI Software} 
Similarly, we sought to identify the dynamically-loaded functions from third party code.
We measured this at a function level, as it is impossible to know which version of this library will be loaded by a particular tool in every deployment.
Table \ref{tab:shared_static_tpl} presents the extent of integration of third-party libraries within open-source VLSI software by listing both shared and statically linked dependencies. 
This table highlights the widespread reliance on external components and record their roles, maintainers and last known update dates. 
Function-level inspection reveals deep integration of shared libraries. Our shared library analysis tool can also identify which specific functions from shared third-party libraries are used. 
Yosys relies on popular libraries like 
    \texttt{libedit} (12 functions), 
    \texttt{libtcl}(24), 
    \texttt{libz}(9), 
    and \texttt{libffi}(7) functions. 
OpenSTA invokes 48 from \texttt{libtcl} and 5 \texttt{libz} functions. 
Further, tools like ABC and Magic utilize \texttt{libedit} (2) and \texttt{libctl} (5), confirming presence of third party libraries across the VLSI design process. Tools such as \texttt{Yosys}, \texttt{OpenSTA}, \texttt{Magic}, and \texttt{OpenTimer} depend on third-party libraries for core functionalities including scripting (\texttt{libtcl8}), command editing (\texttt{libedit}, \texttt{readline}), SAT solving (\texttt{CUDD}, \texttt{MiniSAT}), compression (\texttt{libz}), and graphics (\texttt{libcairo}, \texttt{libXrender}). These transitive and loosely maintained TPLs expand the risk surface by introducing indirect attack paths, particularly in the synthesis and timing analysis stages where silent injection or malformed data can propagate undetected.\\
We also provide a breakdown of the contribution from each third-party library to the execution of each tool.
As shown in Table~\ref{tab:tpl_percentage_all}, a significant portion of these originate from third-party sources. 
Tools like \texttt{Yosys} and \texttt{OpenLane} exhibit over 50\% usage of shared third-party libraries, with \texttt{Yosys} having over 90\% of its binary code originating from statically linked libraries such as \texttt{abc}, \texttt{cudd}, and \texttt{solverl}, indicating a growing reliance on external code that increases SWSC risk. \\
\begin{tcolorbox}[
  colback=green!5,
  colframe=green!60!black,
  title=RQ 1.3 Key Findings,
  fonttitle=\bfseries,
  boxrule=0.5pt,
  arc=2pt,
  left=4pt,
  right=4pt,
  top=2pt,
  bottom=2pt,
  enhanced,
  sharp corners,
  fontupper=\fontsize{8pt}{9pt}\selectfont
] 
Our analysis shows that VLSI software suits relies heavily on third-party/open-source libraries. Across the analyzed software suites, these libraries form a substantial portion of the dependency base (both dynamically and statically linked libraries). Further, several of these libraries correspond to legacy versions last updated over decade ago, indicating potential security and maintenance gaps. The CVE distribution across third-party libraries demonstrates a significant risk surface.

\end{tcolorbox}

Across five major VLSI software, third-party libraries account for up to 62\% of total dependencies, with \texttt{Yosys} and \texttt{OpenLane} leading in external reliance. In \texttt{Yosys}, over 90\% of binary functions in \texttt{yosys-abc} originate from statically linked TPLs. Similarly, \texttt{OpenSTA}'s \texttt{sta} binary shows notable dependence on libraries like \texttt{CUDD} and \texttt{Eigen}. Despite this heavy integration, certain third-party libraries used by VLSI software (e.g., \texttt{libedit}, \texttt{libuuid}, and statically linked components such as \texttt{MiniSAT}) correspond to legacy versions last updated over a decade ago, indicating potential security and maintenance gaps. The CVE distribution across third-party libraries in tools like Yosys (552 contributors, 9 CVEs at 9.8) and OpenLane (722 contributors, 8 CVEs at 9.8) demonstrates a significant risk surface. 

\begin{table*}[t]
\centering
\footnotesize
\renewcommand{\arraystretch}{1.05}
\caption{Third-Party Library Risk Metadata Across VLSI Software Suites}
\label{tab:shared_static_tpl}
\begin{adjustbox}{max width=\textwidth}
\begin{tabular}{|p{1.4cm}|p{3.9cm}|p{1.2cm}|p{1.3cm}|p{3.5cm}|p{1.3cm}|p{1.2cm}|p{1.9cm}|}
\hline
\textbf{VLSI Software} & \textbf{Shared Library Category} & \textbf{Shared: Maintainers} & \textbf{Shared: Oldest Update} & \textbf{Static Library Category} & \textbf{Static: Maintainers} & \textbf{Static: Oldest Update} & \textbf{High-Risk CVEs (\textgreater7.5)} \\
\hline

ABC & Command Editing & 1 & 2012 & SAT/Logic Solving & 1 & 2016 & 0 \\
\hline

Yosys & Command Editing, Scripting, Compression, Interop/Linking, Visualization, UI, System Utils & 552 & 2012 &
Command Editing, Scripting, Parsing, Hashing/Security, SAT/Logic Solving, Interop/Linking, Math, File Format, Hierarchical Analysis
& 881 & 2010 &
9 (9.8), 4 (8.8), 2 (9.1), 1 (9.0), 1 (8.5), 1 (8.4), 1 (8.1), 6 (7.8) \\
\hline

OpenTimer & Scripting & 94 & 2022 & Parsing & 1 & 2018 & 0 \\
\hline

OpenLane & Scripting, Compression, Graphics, Parsing, X11/Display, System Utils & 722 & 2010 &
n/a & n/a & n/a & 8 (9.8), 4 (8.8) \\
\hline

OpenSTA & Scripting, Compression & 156 & 2022 &
Graphics, SAT/Logic Solving, Math & 572 & 2015 &
6 (9.8), 2 (8.8), 3 (7.8) \\
\hline

Magic & Graphics, Scripting & 436 & 2012 &
Scripting, Graphics & 436 & 2012 & 0 \\
\hline

\end{tabular}
\end{adjustbox}
\end{table*}

\definecolor{DetectBlue}{RGB}{0,102,204}

\begin{table}[t]
\centering
\footnotesize
\caption{Third-Party Library Usage in VLSI Software Suites}
\label{tab:tpl_percentage_all}
\begin{tabular}{|l|l|l|}
\hline
\textbf{VLSI} & \textbf{Shared}    & \textbf{Static TPL} \\
\textbf{Tool} & \textbf{Libraries} &  \textbf{(function \%)} \\
\hline
abc         & 16.67\% &  0.00\%    \\
yosys       & 62.00\% & 90.30\%  \\
OpenTimer   & 12.50\% & 0.17\%   \\
OpenLane    & 57.80\% & 0.00\%    \\
OpenSTA     & 11.10\% & 5.37\%   \\
\hline
\end{tabular}
\end{table}

\begin{figure}[t]
    \centering
    \captionsetup{justification=raggedright,singlelinecheck=false}
    \includegraphics[width=0.89\columnwidth]{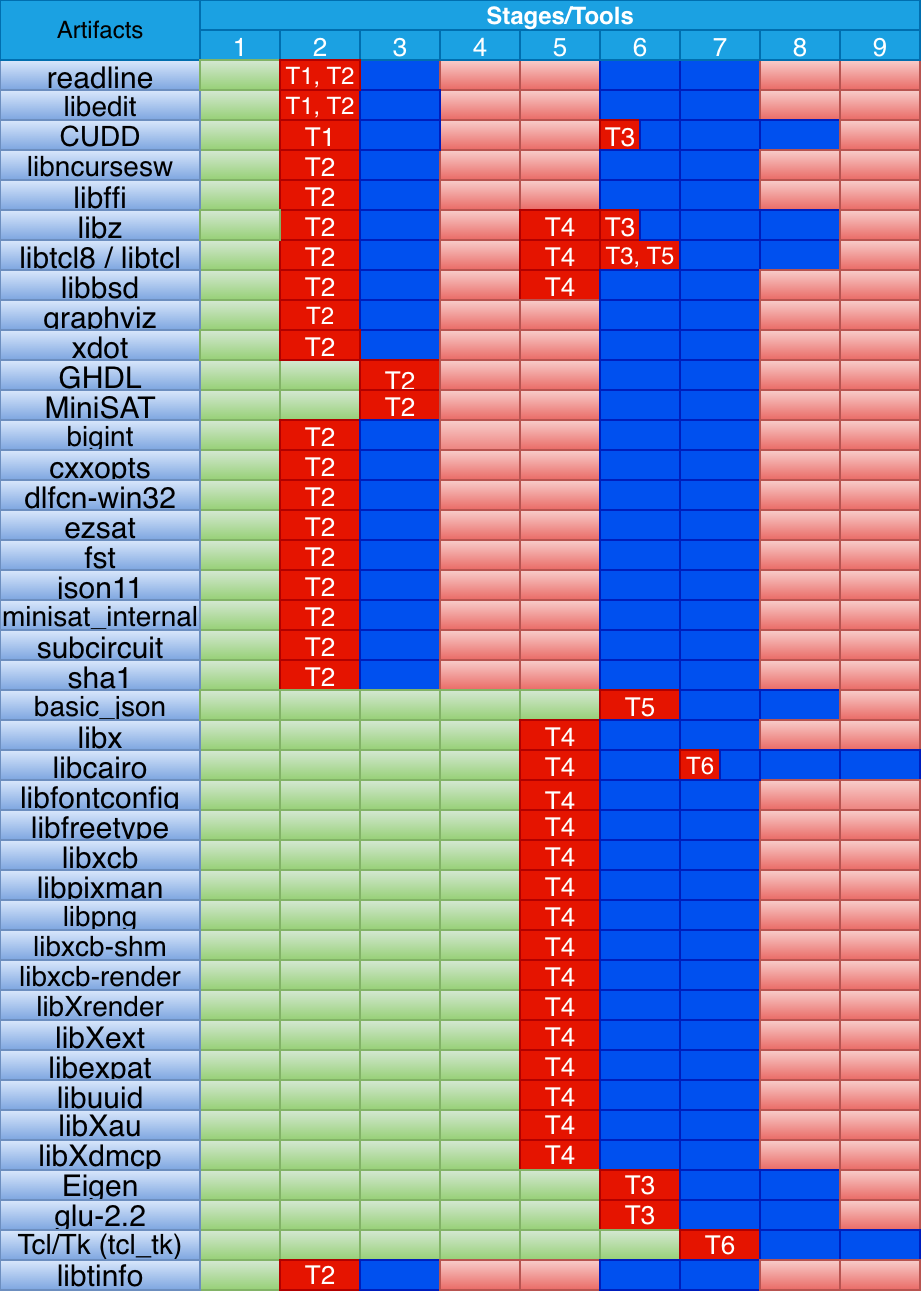}
\caption{\textbf{Library Exposure Across VLSI Flow Stages.}
Stages correspond to the VLSI design steps shown in Figure~\ref{step}.
\protect\textcolor{AffectLightRed}{\rule{6pt}{6pt}}~Cells mark stages affected by a
vulnerable library within a tool (T1--T6), while
\protect\textcolor{StageGreen}{\rule{6pt}{6pt}}~cells denote unaffected stages.
\protect\textcolor{ToolRed}{\rule{6pt}{6pt}}~Tool-specific exposure is labeled by
tool identifiers (T1--T6).
Tools are: T1=ABC, T2=Yosys, T3=OpenSTA, T4=OpenLane, T5=OpenTimer,
and T6=Magic.
\protect\textcolor{DetectBlue}{\rule{6pt}{6pt}}~Annotations show downstream detection phases
(Table~\ref{tab:detection_rubric}).}
\label{fig:library_exposure}
\end{figure}

Further, Fig.~\ref{fig:library_exposure} shows the exposure of third-party libraries across the VLSI design flow. Each row lists a library associated with a specific tool, while columns represent design stages. The colored cells mark the stages where a given library is used, revealing its span and persistence within the flow.

These empirical observations provide the evidence used in our feasibility analysis of the attacks identified by the VLSI-SWSC attack taxonomy(section: \ref{sec:outcomes}).

\section{VLSI SWSC Attack Taxonomy}
\label{sec:outcomes}

Before answering \textbf{RQ2} we must develop an end-to-end model that accounts for software-supply-chain attack vectors, \(A_{\mathrm{sw}}\), that enable capabilities, \(M_{\mathrm{sw}}\), that can lead to a VLSI design attack, \(A_{\mathrm{vlsi}}\), at a particular stage to achieve an outcome \(O\).
This allows us to identify exhaustively all \emph{possible} attacks.
Further, we reduce the list of possible attacks with a second pass to gauge \emph{feasibility} by leveraging risk indicators from the section above.



Formally, our modeled attack paths follow the ensuing sequence on the software side:

\[
I_{\mathrm{sw}} \rightarrow P_{\mathrm{sw}} \rightarrow E_{\mathrm{sw}},
\]

where $I_{\mathrm{sw}}$ denotes the SWSC injection or compromise point, $P_{\mathrm{sw}}$ denotes propagation through the software supply chain, and $E_{\mathrm{sw}}$ denotes exploitation when the compromised software element is consumed. 
We use this three-stage sequence as the conceptual basis for characterizing software supply-chain attacks.

In the synthesized taxonomy, $A_{\mathrm{sw}}$ denotes a software supply-chain attack category that forms the first element of the VLSI-SWSC attack path. Each software supply-chain attack category $A_{\mathrm{sw}}$ can enable a software-side capability gained by the attacker, denoted $M_{\mathrm{sw}}$. Here, $A_{\mathrm{sw}}$ captures how the attacker affects the software supply chain, while $M_{\mathrm{sw}}$ captures the software-side capability enabled by that attack.

The software-side capability $M_{\mathrm{sw}}$ is then mapped to a VLSI-side attack class $A_{\mathrm{vlsi}}$, which captures what the attacker can cause within the VLSI design flow. The VLSI-side attack class can occur at a design stage $S$ and affect a security outcome $O$. Thus, the complete operational mapping of the model is

\[
A_{\mathrm{sw}}
\rightarrow M_{\mathrm{sw}}
\rightarrow A_{\mathrm{vlsi}}
\rightarrow S
\rightarrow O.
\]

The connection between the SWSC definition and the VLSI-side realization occurs through $M_{\mathrm{sw}}$. The exploitation stage $E_{\mathrm{sw}}$ provides the attacker with a software-side capability; this capability is represented by $M_{\mathrm{sw}}$ and determines which VLSI-side attack class $A_{\mathrm{vlsi}}$ can subsequently be realized. Therefore, the conceptual SWSC sequence and the operational taxonomy are connected as

\[
I_{\mathrm{sw}} \rightarrow P_{\mathrm{sw}} \rightarrow E_{\mathrm{sw}}
\quad \Rightarrow \quad
A_{\mathrm{sw}} \rightarrow M_{\mathrm{sw}}
\rightarrow A_{\mathrm{vlsi}} \rightarrow S \rightarrow O.
\]

We define a complete VLSI-SWSC attack path as

\[
A_p =
\langle A_{\mathrm{sw}}, M_{\mathrm{sw}},
A_{\mathrm{vlsi}}, S, O \rangle,
\]

where $A_p$ denotes the attack path, $A_{\mathrm{sw}}$ denotes the SWSC attack category, $M_{\mathrm{sw}}$ denotes the software-side attacker capability enabled by that category, $A_{\mathrm{vlsi}}$ denotes the corresponding VLSI-side attack class, $S$ denotes the VLSI design stage where the attack can be realized or propagated, and $O$ denotes the resulting security outcome. The complete category names and definitions for $A_{\mathrm{sw}}$, $M_{\mathrm{sw}}$, $A_{\mathrm{vlsi}}$, $S$, and $O$ are provided in Appendix~A.

The derived model is shown in Fig. \ref{model} and details are presented as follows. 

\begin{figure}[htbp]
  \centering
\includegraphics[width=0.95\linewidth, trim=80 0 67 0]
{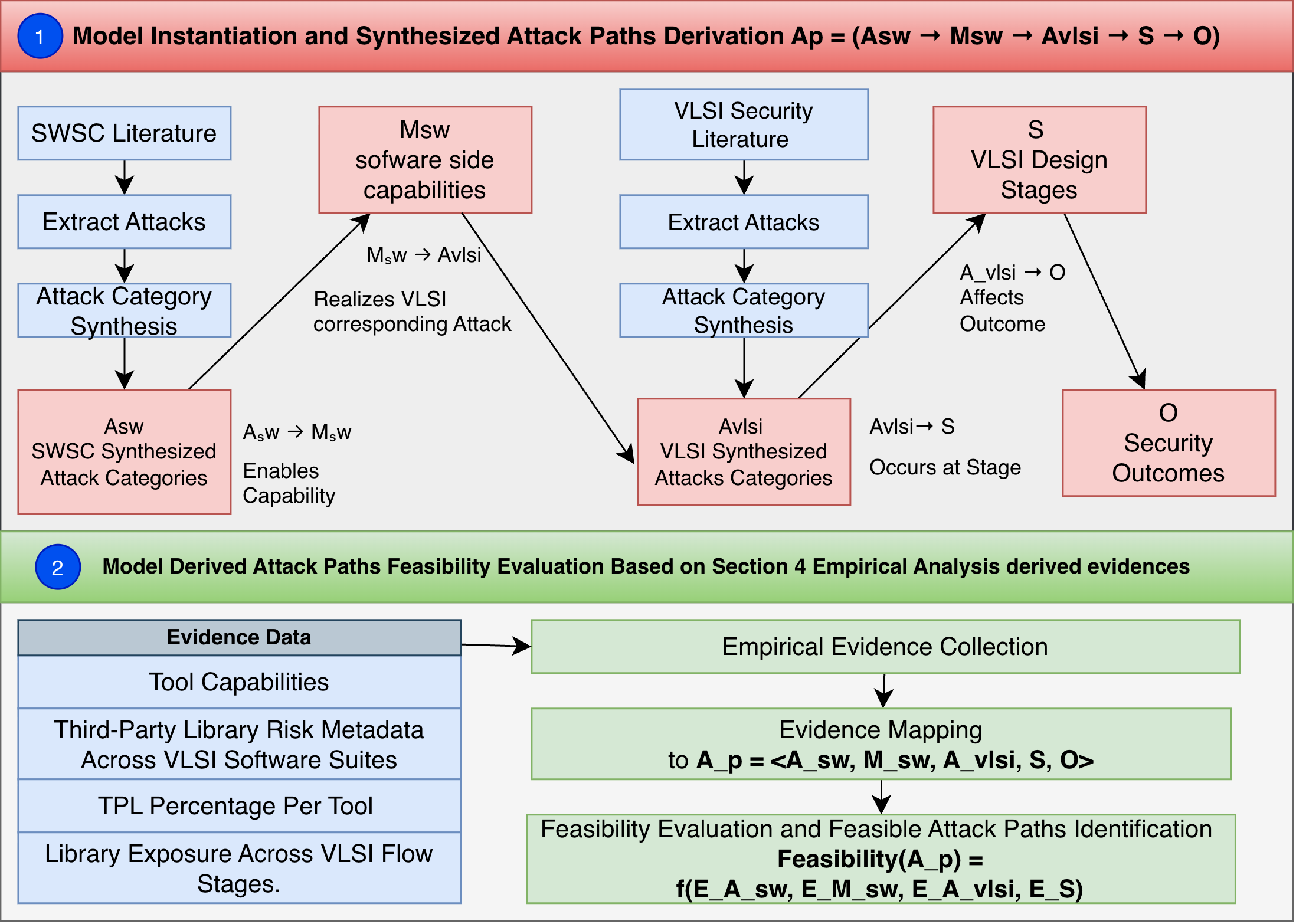}
\caption{Overview of the VLSI-SWSC model. The first panel derives attack paths from SWSC and VLSI-security literature, while the second evaluates their feasibility using empirical evidence from Section~4.}
  \label{model}
\end{figure}

\paragraph{Attack Path Exploration:}
Given the attack-path model defined above, we synthesize VLSI-SWSC attack paths by composing valid relationships between consecutive model elements. Rather than constructing an unrestricted Cartesian product, a path is generated only when an $A_{\mathrm{sw}}$ category enables an $M_{\mathrm{sw}}$ capability, that capability can be realized as an $A_{\mathrm{vlsi}}$ attack class, and the corresponding VLSI-side attack can occur at a design stage and affect a security outcome.

The final attack-path synthesis rule implemented in our model is:

\begin{footnotesize}
\begin{verbatim}
attack_path(A_sw, M_sw, A_vlsi, Stage, Outcome) :-
    valid_asw_msw_pair(A_sw, M_sw),
    valid_msw_avlsi_pair(M_sw, A_vlsi),
    valid_causes(A_vlsi, Stage, Outcome).
\end{verbatim}
\end{footnotesize}
We constructed the Prolog model by synthesizing attack categories from prior software supply-chain and VLSI-security literature. Specifically, we encoded software supply-chain attack types ($A_{\mathrm{sw}}$), software-side attacker capabilities ($M_{\mathrm{sw}}$), VLSI-side attack classes ($A_{\mathrm{vlsi}}$), design-flow stages ($S$), and security outcomes ($O$) as model entities. We then added reviewed mappings between these entities. The $A_{\mathrm{sw}}\rightarrow M_{\mathrm{sw}}$ mappings capture how software supply-chain attack types can lead to attacker capabilities in the software environment, while the $M_{\mathrm{sw}}\rightarrow A_{\mathrm{vlsi}}$ mappings capture how those capabilities can realize VLSI-side attack classes. Finally, the $A_{\mathrm{vlsi}}\rightarrow S$ and $A_{\mathrm{vlsi}}\rightarrow O$ mappings capture where each VLSI-side attack class may occur and which security property it may affect. Together, these mappings define possible paths of the form $A_{\mathrm{sw}}\rightarrow M_{\mathrm{sw}}\rightarrow A_{\mathrm{vlsi}}\rightarrow S\rightarrow O$.

Here, \texttt{valid\_asw\_msw\_pair} requires that $A_{\mathrm{sw}}$ enables $M_{\mathrm{sw}}$; \texttt{valid\_msw\_avlsi\_pair} requires that $M_{\mathrm{sw}}$ can realize $A_{\mathrm{vlsi}}$; and \texttt{valid\_causes} requires that $A_{\mathrm{vlsi}}$ is possible at stage $S$ and affects outcome $O$. These constraints determine which combinations form valid model-generated VLSI-SWSC attack paths. 

\paragraph{Feasibility Evaluation:}
We treat the generated paths as \emph{possible} model paths rather than \emph{plausible} attacks. To assess their feasibility, we add a separate evidence layer tied to the same model categories. We manually extract empirical evidence from Table~\ref{tab:tool_system_capabilities},
Table~\ref{tab:shared_static_tpl}, Table~\ref{tab:tpl_percentage_all}, and
Figure~\ref{fig:library_exposure}, including tool behavior, third-party tool
invocation, shell-injection exposure, dependency exposure, CVEs, third-party
library percentages, static-linking evidence, and library exposure across stages. We then map this evidence back to the existing $A_{\mathrm{sw}}$, $M_{\mathrm{sw}}$, and $A_{\mathrm{vlsi}}$ categories. This keeps the feasibility analysis connected to the Prolog model rather than introducing a separate scoring vocabulary, and bridges the model-generated path space with the empirical observations from our tool analysis.
The second pass evaluates the feasibility of the synthesized attack paths using evidence collected through the empirical analysis in Section~\ref{sec:software suites}. We encode the empirical observations as evidence facts and map them to the corresponding $A_{\mathrm{sw}}$, $M_{\mathrm{sw}}$, and $A_{\mathrm{vlsi}}$ categories, while separately recording the VLSI design stages observed in our analysis.  At the category level, \emph{high} denotes strong empirical support, \emph{medium} denotes partial or indirect support, and \emph{low} denotes insufficient support under the current evidence set. The complete evidence mappings and feasibility rubric are provided in Appendix~A.

The principal rule used to identify high-feasibility attack paths is:

\begin{footnotesize}
\begin{verbatim}
high_feasibility_path(A_sw, M_sw, A_vlsi,
                      Stage, Outcome) :-
    attack_path(A_sw, M_sw, A_vlsi, Stage, Outcome),
    category_high_with_evidence(asw, A_sw),
    category_high_with_evidence(msw, M_sw),
    category_supported_with_evidence(avlsi, A_vlsi),
    observed_stage(Stage).
\end{verbatim}
\end{footnotesize}

Accordingly, a synthesized path is classified as \emph{high feasibility} only when both $A_{\mathrm{sw}}$ and $M_{\mathrm{sw}}$ have high support backed by mapped empirical evidence, $A_{\mathrm{vlsi}}$ has high or medium support backed by mapped evidence, and the corresponding design stage was observed in our empirical analysis.

Paths that do not satisfy the high-feasibility rule are classified as \emph{medium feasibility} when partial support remains across the modeled components. Specifically, medium feasibility requires either support for both $A_{\mathrm{sw}}$ and $M_{\mathrm{sw}}$, high support for either software-side component together with an observed stage, or support for $A_{\mathrm{vlsi}}$ together with an observed stage and support for at least one software-side component. A synthesized path that satisfies neither the high- nor medium-feasibility conditions is classified as \emph{low feasibility}. Thus, low feasibility indicates insufficient support under the current empirical evidence rather than impossibility of the attack. The complete medium- and low-feasibility rules are provided in Appendix~A.

\section{Synthethized Threat Model}
\label{sec:ThreatModel}
Having described the abstract model to evaluate attacks, we instantiate the model using existing data and research to answer RQ2.
\subsection{Model Instantiation}

Having described the abstract model to evaluate attacks, we input data from the literature to the model to answer RQ2.

We identified classes from literature for all of $A_{\mathrm{sw}}$, $M_{\mathrm{sw}}$, $A_{\mathrm{vlsi}}$, $S$, and $O$ to instantiate a model using peer-reviewed attack information.
This allows us to build attack paths from well-established data that can ultimately aid in answering RQ2.
Table~\ref{tab:model_instantiation_sources} outlines the sources and the types of information provided. The extracted vectors were normalized into 7 $A_{\mathrm{sw}}$ categories and 8 $A_{\mathrm{vlsi}}$ classes; the associated software-side mechanisms were consolidated into 8 $M_{\mathrm{sw}}$ capability classes.

\begin{table}[htpb]
\footnotesize
\centering
\caption{Sources used to instantiate the VLSI-SWSC model.}
\label{tab:model_instantiation_sources}
\begin{tabular}{|c | c | c | c|}
\hline
{\bf Source} & {\bf Reference} & {\bf Data Type} & {\bf Class Count} \\
\hline
\hline
Ohm et al. & \cite{ohm2020backstabber} & SWSC attack vectors & 17 \\
\hline
Gokkaya et al. & \cite{gokkaya2026software} & SWSC attack vectors & 15 \\
\hline
Gu et al. & \cite{gu2023investigating} & SWSC attack vectors & 12 \\
\hline
Ladisa et al. & \cite{ladisa2023sok} & SWSC attack vectors & 51 \\
\hline
Liu et al. & \cite{liu2016vlsi} & VLSI attack vectors & 15 \\
\hline
Pundir et al. & \cite{pundir2022analyzing} & VLSI attack vectors & 20 \\
\hline
Tehranipoor et al. & \cite{tehranipoor2024secure} & VLSI attack vectors & 8 \\
\hline
Tehranipoor et al. & \cite{tehranipoor2010survey} & VLSI attack vectors & 26 \\
\hline
Xiao et al. & \cite{xiao2016security} & VLSI attack vectors & 11 \\
\hline
\end{tabular}
\label{tab:model_instantiation_sources}
\end{table}
We utilize the evidence of Section~\ref{sec:software suites} to enrich the feasibility analysis. The resulting evidence set contains 50 evidence facts and 40 evidence-to-category mappings.

\subsection{Model Derived Attack Paths}

Applying these rules to the 717 synthesized attack paths results in \textbf{60 high-feasibility attacks}, \textbf{346 medium-feasibility attacks}, and \textbf{311 low-feasibility attacks}. These classifications quantify the degree of empirical support for model-generated possible attacks and do not constitute experimental validation of all identified attacks.

Although hardware Trojan/backdoor attacks form the largest class in the synthesized space with 216 paths, the 60 high-feasibility paths are concentrated in access abuse/system disruption (32 paths) and design tampering/malicious modification (28 paths). All 60 originate from build/toolchain compromise or vulnerability/configuration exploitation and span multiple VLSI design stages from HLS through physical verification and signoff. The high-feasibility paths affect integrity and availability equally, with 30 paths each.

The complete attack-category definitions, evidence mappings, feasibility rubric, and the 60 high-feasibility paths are provided in Appendix~\ref{s-codes}. The complete set of 717 synthesized attack paths and their classifications is available in the accompanying Open Science artifact.

\section{Attack \& Defense Analysis}
\label{sec:feasibility_defense}

This section presents a systematic evaluation of our software supply chain attacks to understand \textit{how} effectively existing defenses mitigate them across the VLSI design process (see Appendix~\ref{app:scode-legend} for full rubric definitions).

To evaluate the coverage of existing defenses against VLSI SWSC attacks, we adopt a systematized mapping strategy grounded in the CIS software supply chain security guide. We align each CIS-based mitigation domain with published academic and industry techniques, grouped and encoded as strategy codes (S1–S147).This section walks through each mitigation category row by row, identifying relevant S-codes, their supporting literature, and the rationale for their suitability against the proposed VLSI SWSC attack vector. Table \ref{tab:defense_mapping_cis} shows the CIS based mitigation mapping across VLSI SWSC attacks.
\paragraph{Source Code:}The Source Code layer encompasses techniques targeting developer actions, repository configuration, contributor access, third-party inclusion, and static code risk detection. Multiple strategies (S1–S5) from prior work (see Appendix) attempt to mitigate supply chain threats at this stage.\\
\textbf{Code Changes (S1):}
Techniques such as commit signature enforcement, threshold approvals, and reviewer bypass prevention~\cite{yelgundhalli2025rethinking, holtgrave2025attributing, kalu2024industry, duan2020towards, ohm2020backstabber, akinsola2024enhancing} aim to block unauthorized or stealthy code insertions. These strategies directly address $A_{p,11}$ and partially address $A_{p,1}$, $A_{p,2}$, $A_{p,3}$, and $A_{p,4}$ by hardening the commit workflow. However, they offer minimal protection against $A_{p,13}$, $A_{p,14}$, $A_{p,15}$, and $A_{p,16}$ or $A_{p,8}$, $A_{p,9}$, and $A_{p,10}$, which may occur post-merge through tool misuse or flawed design artifacts. See S1 in Appendix.\\ 
\textbf{Repository Management (S2):}
Trust-enhancing mechanisms such as sigstore signing, per-package repository delegation, and role separation~\cite{moore2023artemis, newman2022sigstore, nocera2023software} help verify repository authenticity and mitigate takeover risks. These efforts contribute to $A_{p,11}$ resilience but lack runtime enforcement, leaving $A_{p,12}$ and $A_{p,13}$, $A_{p,14}$, $A_{p,15}$, and $A_{p,16}$ under-addressed. See S2 in Appendix.\\
\textbf{Contribution Access (S3):}
CI-based access control, onboarding filters, and change logging~\cite{cui2024role, bhattacharyya2023study, wermke2023always} improve contributor vetting and traceability, relevant to $A_{p,11}$ and $A_{p,12}$. Still, these rely on pre-deployment enforcement and offer limited reach into $A_{p,8}$, $A_{p,9}$, and $A_{p,10}$ (faulty or unavailable designs) or $A_{p,13}$ and $A_{p,14}$ (tool disruption), which involve unintended behaviors post-contribution. See S3 in Appendix.\\ 
\textbf{Third Party (S4):}
Community-scale protections like SBOM mandates and cross-project trust policies~\cite{wermke2022committed, williams2024psscrm} help govern external code inclusion. These mitigate $A_{p,11}$ and $A_{p,12}$ when third-party components are involved but do not explicitly trace downstream runtime failures or dynamic misbehavior, such as $A_{p,8}$, $A_{p,9}$, $A_{p,10}$, $A_{p,13}$, and $A_{p,14}$. See S4 in Appendix.\\
\textbf{Code Risks (S5):}
LLM-driven tools for vulnerability discovery and malware pattern detection~\cite{ashiwal2024llm, zahan2024shifting} target latent flaws in committed code. While promising for identifying security weaknesses that may enable information leakage or malicious design modification, these methods lack guarantees for timing-based exfiltration or environment-specific failures, thus providing only partial mitigation for information leakage, design tampering, or design-flow disruption. See S5 in Appendix.  
\paragraph{Build Pipelines}
The Build Pipelines layer secures compilation and integration workflows. Strategies S6--S9 target reproducibility, build isolation, and CI/CD hardening (Appendix). \\
\textbf{Build Environment (S6):}
Reproducible builds and deterministic patching~\cite{fourne2023flossing, butler2023reproducible, ren2022patching} defend against $A_{p,11}$ and reduce $A_{p,1}$--$A_{p,4}$ risks, but do not detect $A_{p,12}$ or $A_{p,8}$--$A_{p,10}$, nor address runtime degradation in $A_{p,13}$--$A_{p,16}$.\\
\textbf{Build Worker (S7):}
Worker isolation and sandboxing~\cite{enck2022topfive} limit attacker persistence. While effective against build-time tampering ($A_{p,11}$), they do not monitor generated artifacts for malicious logic or design modification ($A_{p,1}$, $A_{p,2}$, $A_{p,3}$, and $A_{p,4}$) or compromised flow/build-script behavior ($A_{p,12}$).\\
\textbf{Pipeline Instructions (S8):}
Scripting hardening and CI/CD control flows~\cite{tamanna2024slsa, chandramouli2024nist} help disrupt attacker-triggered builds ($A_{p,11}$ and $A_{p,12}$), but lack runtime or testing enforcement, thus under-addressing $A_{p,8}$, $A_{p,9}$, and $A_{p,10}$ (faulty or unavailable designs) as well as $A_{p,13}$, $A_{p,14}$, $A_{p,15}$, and $A_{p,16}$ (tool or flow disruption).\\
\textbf{Pipeline Integrity (S9):}
Policy-as-code and artifact logging~\cite{enck2023s3c2gov, tran2022s3c2sep} strengthen build auditability. While for traceability, these strategies are retrospective and insufficient for dynamic threats like testing flaws ($A_{p,8}$, $A_{p,9}$, and $A_{p,10}$) or performance and flow sabotage ($A_{p,15}$ and $A_{p,16}$).
\paragraph{Dependencies}
The Dependencies layer mitigates risks from external packages via detection, validation, and hygiene enforcement (Appendix).\\
\textbf{Third Party Packages (S10, S15, S19):}
Static diffing, behavioral modeling, and typo-squatting detection~\cite{ladisa2023hitchhiker, froh2023differential, taylor2020spellbound} address malicious or misused dependencies ($A_{p,11}$, $A_{p,1}$, $A_{p,2}$, $A_{p,3}$, and $A_{p,4}$) through structural and naming anomaly detection. However, these strategies often rely on known patterns or blacklists, offering limited protection against novel runtime exfiltration scripts ($A_{p,12}$) or latent availability degraders ($A_{p,13}$, $A_{p,14}$, $A_{p,15}$, and $A_{p,16}$).\\
\textbf{Validate Packages (S11–S18):}
Advisory mapping~\cite{nguyen2022vulcurator, dunlap2023vfcfinder}, silent patch tracking~\cite{nguyen2023vffinder}, and freshness scoring~\cite{zahan2023scorecard, renovatebot2024renovate} verify dependency safety and reduce legacy or residual information leakage, design vulnerabilities that may enable malicious design modificationmpering, or design-flow disruption, but do not prevent malicious code injection or intentional flow degradation. CVE-based tools may miss backdoored yet functional packages. Overall, dependency validation improves hygiene but lacks semantic inspection of functional intent, making it partially effective against integrity and minimally effective against confidentiality or availability threats.

\paragraph{Artifacts}
This category focuses on verifying, tracking, and restricting access to artifacts to preserve integrity and traceability.\\
\textbf{Verification (S20)}
Tools Memptec~\cite{halder2024memptec} leverage anomaly detection to flag malicious packages. While effective against backdoored tools ($A_{p,11}$) or packaging-based obfuscation, they cannot analyze code intent, limiting their efficacy against flow scripts ($A_{p,12}$) or availability threats ($A_{p,15}$ and $A_{p,16}$).\\
\textbf{Access to Artifacts (S21), Package Registries (S22), Origin Traceability (S23).}
These strategies enhance provenance and access control. SLSA-aligned artifact forensics~\cite{ohm2020towards} and public registry signing~\cite{schorlemmer2024signing} provide origin authentication and tamper resistance ($A_{p,11}$, $A_{p,1}$, $A_{p,2}$, $A_{p,3}$, and $A_{p,4}$). Argo-SLSA pipelines ~\cite{thariq2025argo} further bind build processes to verified sources. However, they offer little protection once a verified artifact is internally compromised or misused in a flow script ($A_{p,12}$, $A_{p,13}$, and $A_{p,14}$).\\

\begin{table*}[ht]
\centering
\caption{CIS-Based Mitigation Mapping Across VLSI Software Supply Chain Threats.
\textbf{Symbols:} \FM = Fully Met, \PM = Partially Met, \NM = Not Met.
\textbf{Attack-Path Classes:}
$A_{p,1}$ = \{P614, P616\};
$A_{p,2}$ = \{P334, P336, P338\};
$A_{p,3}$ = \{P618, P620, P622\};
$A_{p,4}$ = \{P340, P342, P344, P346\};
$A_{p,5}$ = \{P615, P617\};
$A_{p,6}$ = \{P423, P425\};
$A_{p,7}$ = \{P427, P429, P431, P435\};
$A_{p,8}$ = \{P420, P422, P424, P426\};
$A_{p,9}$ = \{P619, P621, P623, P625, P627\};
$A_{p,10}$ = \{P428, P430, P432, P436\};
$A_{p,11}$ = \{P333, P335, P337, P339\};
$A_{p,12}$ = \{P341, P343, P345\};
$A_{p,13}$ = \{P700, P702, P704, P706\};
$A_{p,14}$ = \{P419, P421\};
$A_{p,15}$ = \{P430, P432, P436\};
$A_{p,16}$ = \{P701, P703, P705, P707, P708, P709, P710, P711,
P712, P713, P716, P717, P624, P626\}.
\textbf{Full Attack Paths:} and
\textbf{S-Codes:} Full legend (S1--S147) See Appendix~\ref{app:scode-legend}.}
\label{tab:defense_mapping_cis}
\vspace{1mm}
\footnotesize
\renewcommand{\arraystretch}{1.15}
\setlength{\tabcolsep}{1.5pt}

\begin{adjustbox}{max width=\textwidth}
\begin{tabularx}{\textwidth}{|p{2.5cm}|p{3.8cm}|p{2.6cm}|*{16}{>{\centering\arraybackslash}X|}}
\hline
\textbf{CIS Technique} & \textbf{Sample References} & \textbf{Strategy (S Codes)} &
\rotatebox{90}{$A_{p,1}$} &
\rotatebox{90}{$A_{p,2}$} &
\rotatebox{90}{$A_{p,3}$} &
\rotatebox{90}{$A_{p,4}$} &
\rotatebox{90}{$A_{p,5}$} &
\rotatebox{90}{$A_{p,6}$} &
\rotatebox{90}{$A_{p,7}$} &
\rotatebox{90}{$A_{p,8}$} &
\rotatebox{90}{$A_{p,9}$} &
\rotatebox{90}{$A_{p,10}$} &
\rotatebox{90}{$A_{p,11}$} &
\rotatebox{90}{$A_{p,12}$} &
\rotatebox{90}{$A_{p,13}$} &
\rotatebox{90}{$A_{p,14}$} &
\rotatebox{90}{$A_{p,15}$} &
\rotatebox{90}{$A_{p,16}$} \\
\hline

Source Code &
\cite{yelgundhalli2025rethinking, holtgrave2025attributing, kalu2024industry, duan2020towards, ladisa2023sok, ohm2020backstabber, al2021security, akinsola2024enhancing, przymus2025wolves, shen2025understanding, wang2025large} &
S1--S11 &
\PM & \PM & \PM & \PM & \PM & \NM & \PM & \NM & \NM & \NM & \NM & \NM & \NM & \NM & \NM & \NM \\

\hline

Repository Management &
\cite{moore2023artemis, newman2022sigstore, nocera2023software} &
S12--S27 &
\PM & \PM & \PM & \NM & \FM & \FM & \PM & \PM & \PM & \NM & \FM & \PM & \PM & \PM & \PM & \PM \\

\hline

Contribution Access &
\cite{cui2024role, bhattacharyya2023study, wermke2023always, holtgrave2025attributing} &
S28--S50 &
\PM & \PM & \PM & \PM & \PM & \PM & \PM & \PM & \PM & \PM & \PM & \PM & \PM & \PM & \PM & \NM \\

\hline

Third Party &
\cite{wermke2023always, wermke2022committed, williams2024psscrm} &
S41--S55 &
\PM & \PM & \PM & \PM & \PM & \PM & \PM & \PM & \PM & \PM & \PM & \NM & \NM & \NM & \NM & \NM \\

\hline

Code Risks &
\cite{ashiwal2024llm, boi2024smart, katsadouros2023can, zahan2024shifting, zhang2024prompt, zhang2023how, liu2023software, benedetti2022alice} &
S56--S62 &
\PM & \PM & \PM & \NM & \NM & \NM & \NM & \NM & \NM & \PM & \PM & \NM & \NM & \NM & \NM & \NM \\

\hline

Build Environment &
\cite{fourne2023flossing, butler2023reproducible, ren2022patching} &
S63--S71 &
\FM & \PM & \PM & \PM & \FM & \NM & \NM & \NM & \NM & \NM & \PM & \PM & \NM & \NM & \NM & \NM \\

\hline

Build Worker &
\cite{enck2022topfive} &
S72--S74 &
\PM & \PM & \PM & \PM & \PM & \PM & \PM & \PM & \PM & \PM & \PM & \PM & \NM & \NM & \NM & \NM \\

\hline

Pipeline Instructions &
\cite{tamanna2024slsa, chandramouli2024nist, gu2023continuous, moore2023artemis} &
S75--S85 &
\FM & \FM & \FM & \PM & \FM & \PM & \PM & \PM & \PM & \PM & \PM & \PM & \NM & \NM & \NM & \NM \\

\hline

Pipeline Integrity &
\cite{enck2023s3c2gov, dunlap2023s3c2feb, enck2022topfive, tran2022s3c2sep, tystahl2024s3c2mar} &
S86--S88 &
\PM & \PM & \PM & \PM & \PM & \PM & \PM & \PM & \PM & \PM & \PM & \PM & \PM & \PM & \PM & \PM \\

\hline

Third Party Packages &
\cite{enck2023s3c2gov, barrsmith2022exorcist, froh2023differential, yu2024maltracker, zahan2022weaklinks, zhang2023mavenpersistence} &
S89--S111 &
\PM & \PM & \PM & \PM & \PM & \PM & \PM & \PM & \PM & \PM & \PM & \PM & \NM & \NM & \NM & \NM \\

\hline

Validate Packages &
\cite{alfadel2023npm, csa2024gsd, cox2015freshness, croft2023dataquality, zahan2024shifting, zhang2023malicious} &
S112--S118 &
\PM & \PM & \PM & \PM & \PM & \PM & \PM & \PM & \PM & \PM & \PM & \PM & \PM & \PM & \PM & \PM \\

\hline

Verification &
\cite{halder2024memptec} &
S119--S125 &
\PM & \PM & \PM & \PM & \PM & \PM & \PM & \PM & \PM & \PM & \PM & \PM & \PM & \PM & \PM & \PM \\

\hline

Access to Artifacts &
\cite{ohm2020towards, tamanna2024unraveling} &
S126--S130 &
\PM & \PM & \PM & \PM & \PM & \PM & \PM & \PM & \PM & \PM & \PM & \PM & \PM & \PM & \PM & \PM \\

\hline

Package Registries &
\cite{schorlemmer2024signing} &
S131--S135 &
\PM & \PM & \PM & \PM & \PM & \NM & \NM & \PM & \PM & \PM & \PM & \PM & \PM & \PM & \PM & \PM \\

\hline

Origin Traceability &
\cite{thariq2025argo} &
S136--S140 &
\PM & \PM & \PM & \PM & \PM & \PM & \PM & \PM & \PM & \PM & \PM & \PM & \PM & \PM & \PM & \PM \\

\hline

Deployment Configuration &
\cite{chiari2022iacsurvey, verdet2023iacstudy, saavedra2022glitch, rahman2019smells} &
S141--S144 &
\PM & \NM & \NM & \FM & \PM & \NM & \FM & \NM & \NM & \FM & \PM & \FM & \NM & \PM & \PM & \NM \\

\hline

Deployment Environment &
\cite{harness2022cicd, tripwire2022cd} &
S145--S147 &
\PM & \NM & \NM & \PM & \PM & \NM & \PM & \PM & \PM & \PM & \PM & \PM & \PM & \NM & \PM & \PM \\

\hline
\end{tabularx}
\end{adjustbox}
\end{table*}

\section{Threat Model Evaluation: Case Studies}
\label{sec:casestudy}

Finally, we evaluate our VLSI-SWSC attack taxonomy on a reference implementation of a VLSI design flow. All experiments were conducted in a controlled laboratory environment running Ubuntu 24.04.1 LTS on an x86-64 system equipped with a 13th Gen Intel Core i7-13700K processor and 32 GB of RAM. We implement three attack paths synthesized by our taxonomy model and separately discuss one additional taxonomy-derived scenario as speculative. The following describes each case study and its corresponding synthesized attack path.

\textbf{S1: P376 -- Build/Toolchain Compromise Leading to Information Leakage:}

P376 is characterized by
\(A_{sw}\): build/toolchain compromise,
\(M_{sw}\): build/toolchain execution control,
\(A_{vlsi}\): information leakage/asset exposure,
\(S\): logic synthesis, and
\(O\): confidentiality.

In this case study, we instantiate P376 using Yosys at the logic-synthesis stage. Yosys relies on the third-party \texttt{zlib} library to decompress compressed design files (e.g., \texttt{.v.gz}). We created a malicious version of \texttt{zlib} that hijacks the \texttt{gzread()} function. When Yosys executes with the compromised library, the malicious function intercepts reads of the Verilog design, copies the design data, and writes it to an external file.
The attack therefore exposes the circuit design during logic synthesis, resulting in information leakage and loss of IP confidentiality.\\

\textbf{S2: P703 -- Configuration-Based Flow Disruption Affecting Availability:}

P703 is characterized by
\(A_{sw}\): vulnerability/configuration exploitation,
\(M_{sw}\): vulnerability/configuration-based control,
\(A_{vlsi}\): access abuse/system disruption,
\(S\): logic synthesis, and
\(O\): availability.

In this case study, we instantiate P703 using Yosys at the logic-synthesis stage. Yosys uses the standard C library to perform file operations, including the \texttt{fopen()} function for accessing Verilog input files. We created a malicious shared object that overrides \texttt{fopen()} and loaded it through the \texttt{LD\_PRELOAD} mechanism (e.g., to mimic the malicious dlopen hijack from the xz-utils compromise). The malicious function denies access to the intended RTL file and redirects the tool to a fake or empty input file.

This configuration-based control disrupts synthesis by preventing Yosys from processing the required RTL artifact. As a result, the expected synthesis output is missing or invalid, disrupting subsequent stages of the VLSI design flow and affecting availability.\\
\textbf{S3: P616 -- Configuration-Based Design Tampering Affecting Integrity:}

P616 is characterized by
\(A_{sw}\): vulnerability/configuration exploitation,
\(M_{sw}\): vulnerability/configuration-based control,
\(A_{vlsi}\): design tampering/malicious modification,
\(S\): logic synthesis, and
\(O\): integrity.

In this case study, we instantiate P616 using Yosys at the logic-synthesis stage. We created a malicious shared object that overrides the \texttt{read()} function used during Verilog file processing. When Yosys reads the design input, the malicious function modifies the input stream by inserting an additional line, such as \texttt{invalid\_gate xyz();}, before the legitimate data is processed.

This configuration-based control modifies the design artifact during synthesis, resulting in a malformed netlist containing invalid components. 
The attack therefore tampers with the design and violates its integrity, while also causing downstream timing, placement, and layout stages to fail.\\
\textbf{S4: P624 -- Speculative Configuration-Based Timing Manipulation Affecting Integrity:}

P624 is characterized by
\(A_{sw}\): vulnerability/configuration exploitation,
\(M_{sw}\): vulnerability/configuration-based control,
\(A_{vlsi}\): design tampering/malicious modification,
\(S\): static timing analysis, and
\(O\): integrity.

In this speculative case study, we consider an attack on the static timing analysis stage enabled by third-party library dependencies in VLSI timing tools. Tools such as OpenTimer, OpenSTA, and OpenLane rely on scripting and data-processing libraries that can expose software-side attack surfaces.
An attacker could load a malicious parsing library that intercepts functions responsible for processing timing values. By modifying the returned timing constraints, the compromised library could cause the timing tool to incorrectly accept violated setup or hold constraints as valid.
This configuration-based control could therefore manipulate timing-analysis results, allowing incorrect timing information to propagate through signoff and compromising the design.

\section{Related Work}

The field of Hardware Supply Chain Security has been established in research and development. This area of research focuses on detecting and preventing malicious hardware changes (i.e., Trojans) during manufacturing as well as preventing the theft of IP when outsourcing manufacturing to a third party — who is generally overseas. To avoid malicious tampering in hardware supply chains, various approaches have been proposed.

Hardware designs face increasing threats from side-channel attacks, logic-locking vulnerabilities, fault injection, hardware Trojans, and IP piracy. Recent research focuses on identifying and mitigating these hardware supply-chain vulnerabilities across abstraction levels, from RTL to gate level. \cite{srivastava2024scar}, cross talk \cite{choudhury2024crosstalk}, and electromagnetic leakages in complex computing environments \cite{bepary2025towards}. SCAR \cite{srivastava2024scar} uses graph neural networks on RTL-level control-data flow graphs for pre-silicon identification and mitigation of power side-channel vulnerabilities, achieving high localization accuracy (94.49 \%), with automated code-level mitigations generated by large language models (LLMs). This work \cite{choudhury2024crosstalk} demonstrates a crosstalk-based quantum side channel attack in multi-tenant NISQ environment, using graph learning to identify victim circuits from CNOT gate signatures with 85.7 \% accuracy. Recent studies have also focused on detecting hardware Trojans and evaluating the resilence of logic locking and obfuscation techniques, revealing practical limitations in securing layouts against reverse engineering and SAT based attacks. Additionally, \cite{meng2024needle} transforms IC layouts into graphs and applies graph neural networks to classify functional regions, detecting and localizing hardware trojans based on low confidence predictions without a golden IC. Further, \cite{knechtel2025trojan} presents a large scale red versus blue team evaluation of fabrication-time hardware Trojan insertion, demonstrating that even optimized 7nm IC layouts with strict design constraints remain vulnerable, and releases a comprehensive benchmarking framework for continued research. Moreover, \cite{han2021does} demonstrates that current logic locking techniques based on the corrupt-and-correct (CAC) model are vulnerable to structural attacks enbaled by EDA tools optimizations, introducing the SPI attack which breaks all tested CAC schemes with 100 \% success, calling for co-design of secure locking and EDA flows. Further, \cite{han2020multi} enhances stripped functionality logic locking (SFLL) by proposing multi-objective node selection strategies that improve output corruption and coverage, with MinCn showing superior scalability and effectiveness for larger key sizes. Further, \cite{datta2022security} work investigates clock signal obfuscation as a logic locking primitive showing that while clock based schemes offer limited security per overhead, targeted designs may benefit from specialized de-obfuscation resistance.

Recent studies expose key vulnerabilities in the hardware IP lifecycle, including SAT-based de-obfuscation, insecure provisioning, and unverified protection mechanisms. Delay-locked circuits remain susceptible to TimingSAT attacks, prompting the proposal of a stripped-functionality locking scheme for enhanced resilience \cite{chakraborty2020evaluating}. SAP introduces a lightweight hardware IP to enable secure asset provisioning and SoC authentication in untrusted environments \cite{sami2024sap}. Despite advances in CAD-driven IP protection (e.g., obfuscation, watermarking), scalability and robustness issues persist under globalized supply chains \cite{bhunia2022hardware}. Architectural oversights—not physical tampering—are shown to underlie most PCB-level attacks, emphasizing the role of sound design principles \cite{harrison2024sok}. Logic locking and trust verification continue to be key defenses against Trojans and IP theft \cite{salmani2019special}, while IV\&V frameworks ensure these protections hold under real-world EDA flows \cite{tan2021independent}.

Recent work explores pre-silicon security verification and SoC attack models using formal and learning-based approaches. RTL-ConTest applies RTL-level concolic testing to detect processor- and peripheral-level vulnerabilities in open-source SoCs. \cite{meng2021rtl}, while Re-Pen and RL-TPG employ reinforcement learning to generate optimized test patterns for automated vulnerability discovery under limited design knowledge \cite{al2024re}. FuncTeller reveals that eFPGA redaction does not guarantee confidentiality, recovering IP functionality through black-box queries with 85\% accuracy \cite{han2023functeller}. HUnTer systematically verifies system-level vulnerabilities by linking formal threat models to exploitable instruction sequences on SoC architectures \cite{rajendran2024exploring}. Complementarily, formal RTL fault-injection techniques uncover early-stage vulnerabilities in control logic without relying on gate-level representations \cite{uddin2024case}. Clock gating has also been explored for IP protection through programmable integrated clock gate insertion at RTL~\cite{tehranipoor2024clock}. Finally, studies expose software-based side-channel attacks—using power telemetry interfaces and CPU frequency throttling—on Apple SoCs, enabling AES key recovery from user space and challenging assumptions about isolation in modern processors~\cite{chawla2024uncovering}.

\section{Conclusion}
\vspace{-.5em}
This paper presented a systematic analysis of SWSC vulnerabilities within the VLSI design flow, demonstrating how software-level attacks can compromise hardware security. 
Through an empirical analysis of VLSI software suites, we characterized their software composition, assessed tool interactions, and evaluated their security features and operating privileges. 
Building on these insights, we developed a VLSI-SWSC attack taxonomy that links software-side attacks and capabilities to VLSI-side attacks, design stages, and security outcomes. 
The taxonomy further informed the construction of our threat model, capturing potential security outcomes arising from compromised software components. Finally, we evaluated representative taxonomy-derived attacks on an open-source VLSI design flow by implementing three attack scenarios and separately analyzing one additional speculative scenario.
We hope that, through this, a first bridge into mapping the security posture of VLSI software suites from SWSC attacks is created.
\cleardoublepage
\appendix
\section*{Ethical Considerations}
This section discusses the ethical considerations of analyzing software supply chain vulnerabilities in VLSI design flow. Our work focuses on empirical analysis of VLSI software suites and identified attacks that can be carried out from a software supply chain prespective and as a result create software-only attack taxonomy. We did the evaluation of these software-only attacks in a controlled environment.
\begin{itemize}
    \item \textit{Risks and benefits from identified software supply-chain weaknesses:} Our analysis reveals that untrusted open source software dependencies and tool interactions in VLSI design flows can enable software-only attacks affecting the confidentiality, integrity, and availability of the design process. These risks are clearly defined threat model and demonstrated only in controlled environments using open-source tools and synthetic designs.The primary benefit of this analysis is to expose previously underexplored software-driven risks in VLSI pipelines and to inform prioritization and mitigation by tool developers and practitioners.
    \item \textit{Risks and benefits from the existence of this work:} Thsi work provides a systematic taxonomy and threat model for software supply chain risks in VLSI design flow. While such analyses could be misused, we mitigate this risk by avoiding release of exploit code or weaponized artifacts and by reporting attacks at an abstract level.
    \item \textit{Responsible Disclosure} As part of this work, we have engaged with relevant VLSI software vendors regarding the security issues identified in our analysis. Our goal is to responsibly communicate potentially actionable findings and provide vendors an opportunity to review them before broader disclosure. We avoid releasing sensitive implementation details that could enable misuse and limit public reporting to information necessary to support the scientific findings of this work.
\end{itemize}

\section*{Open Science}
We implemented three end-to-end software supply chain attacks---P376, P703, and P616---using only open-source tools and libraries. In P376, we hijack \texttt{gzread()} in \texttt{libz.so} to leak compressed Verilog designs via dynamic preloading (\texttt{LD\_PRELOAD}), demonstrating build/toolchain compromise leading to information leakage. In P703, we hook \texttt{fopen()} in \texttt{glibc} to silently substitute empty or fake RTL input files, causing configuration-based flow disruption that affects availability. In P616, we use a custom wrapper on \texttt{read()} in \texttt{libc.so} to inject malformed gates, resulting in configuration-based design tampering that affects integrity. All attack implementations, input examples, and reproduction instructions are available in our anonymized GitHub repository: 
\url{https://anonymous.4open.science/r/VLSI-E052}

\bibliographystyle{plainurl}
\bibliography{sample-base}

\section{Appendix}
\label{app:scode-legend}

\begin{table*}[t]
\centering
\footnotesize
\label{D}
\caption{\textbf{Detection Rubric.}
Detection verdict is derived by: 
(1) Access Vector baseline, 
(2) Detection Complexity adjustment, and 
(3) CIA Impact adjustment. Final verdict: $D_{\text{final}} = D_{\text{AV}} \oplus DC \oplus CIA$,
where upgrades/downgrades follow: ●=3, ◑=2, ○=1, –=0.}
\label{tab:detection_rubric}

\begin{adjustbox}{max width=\textwidth}
\begin{tabular}{|p{3cm}|p{1.7cm}|p{4.2cm}|p{4.2cm}|p{4.3cm}|}
\hline
\textbf{Metric} & \textbf{Identifier} & \textbf{Description} & \textbf{Example} & \textbf{Evaluation} \\
\hline

\textbf{No Detection Research} &
D:ND &
No published VLSI-specific detection research or tooling. &
Only conceptual or unrelated discussions; no frameworks/datasets. &
\textbf{Not Met (○)} — Represents a clear open research gap. \\
\hline

\textbf{Detection Complexity: High} &
DC:H &
Requires specialized or non-automated methods (formal equivalence, side-channel analysis). &
Hidden logic in gate-level netlist detectable only via deep formal methods. &
\textbf{Low (○)} — Very hard to detect automatically; requires specialized tools. \\
\hline

\textbf{Detection Complexity: Medium} &
DC:M &
Partial automation; manual triage needed. &
Provenance mismatch between netlist and source commit requiring human validation. &
\textbf{Medium (◑)} — Some automation, but relies on expert analysis. \\
\hline

\textbf{Detection Complexity: Low} &
DC:L &
Fully automated detection using standard CI/CD safeguards. &
CI rejects unsigned binaries using SBOM+signature enforcement. &
\textbf{High (●)} — Easily detectable via standard automated checks. \\
\hline

\multicolumn{5}{|c|}{\textbf{CIA Impacts (Included Inside Table)}} \\
\hline

\textbf{Confidentiality Impact: None} &
C:None &
Research provides no confidentiality protections. &
No encryption, isolation, or access control. &
\textbf{Not Met (○)} — No confidentiality controls present. \\
\hline

\textbf{Confidentiality Impact: Partial} &
C:Partial &
Research partially mitigates leakage but lacks complete protection. &
Limited auditing or partial access restriction. &
\textbf{Partially Met (◑)} — Partial IP/data protection only. \\
\hline

\textbf{Confidentiality Impact: Full} &
C:Full &
Strong confidentiality mechanisms preventing IP/data leakage. &
Access control, encryption, provenance tracking. &
\textbf{Fully Met (●)} — Clear confidentiality enforcement. \\
\hline

\textbf{Integrity Impact: None} &
I:None &
Research does not ensure design/tool/integration integrity. &
Unsigned artifacts, no reproducible builds. &
\textbf{Not Met (○)}. \\
\hline

\textbf{Integrity Impact: Partial} &
I:Partial &
Some integrity checks or audit trails, but not complete. &
Signed commits without build reproducibility. &
\textbf{Partially Met (◑)}. \\
\hline

\textbf{Integrity Impact: Full} &
I:Full &
Full integrity guarantees (signatures, attestation, reproducible builds). &
Formally verified, attested artifacts. &
\textbf{Fully Met (●)}. \\
\hline

\textbf{Availability Impact: None} &
A:None &
No coverage against disruption/DoS. &
No redundancy or recovery. &
\textbf{Not Met (○)}. \\
\hline

\textbf{Availability Impact: Partial} &
A:Partial &
Partial/ reactive resilience after disruption. &
Slow detection or partial failover. &
\textbf{Partially Met (◑)}. \\
\hline

\textbf{Availability Impact: Full} &
A:Full &
Explicit uptime/resilience guarantees. &
Redundancy, sandboxing malicious tools. &
\textbf{Fully Met (●)}. \\
\hline

\end{tabular}
\end{adjustbox}
\end{table*}

\begin{figure*}[htbp]
  \centering
  \includegraphics[width=0.75\textwidth]{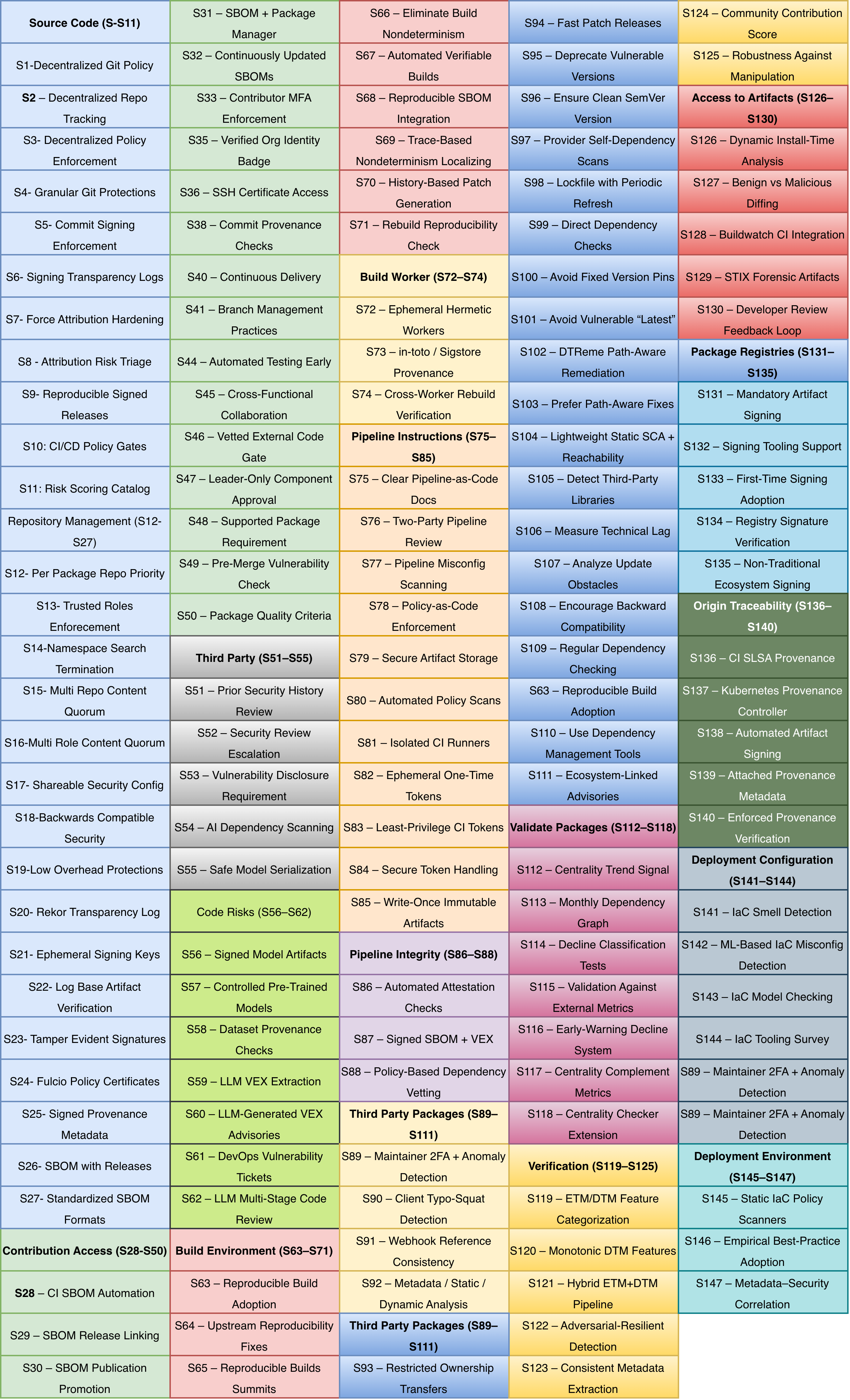}
  \caption{S-codes S1-S147}
  \label{s-codes}
\end{figure*}


\begin{table*}[htbp]
\centering
\footnotesize
\caption{Model entities used in the VLSI-SWSC attack taxonomy.}
\label{tab:model_categories}
\begin{tabular}{p{0.10\textwidth} p{0.34\textwidth} p{0.48\textwidth}}
\hline
\textbf{Element} & \textbf{Category} & \textbf{Definition} \\
\hline
\hline

$A_{\mathrm{sw}}$ &
Package introduction or misselection &
Introduction or selection of an attacker-controlled or misleading software package. \\

$A_{\mathrm{sw}}$ &
Source repository or code tampering &
Unauthorized modification of source code, repositories, or version-control content. \\

$A_{\mathrm{sw}}$ &
Trusted identity or maintainer compromise &
Abuse or compromise of trusted developer, maintainer, or project-level access. \\

$A_{\mathrm{sw}}$ &
Build or toolchain compromise &
Compromise of build processes, build infrastructure, compilers, or development tools. \\

$A_{\mathrm{sw}}$ &
Distribution, hosting, or update compromise &
Manipulation of software distribution, hosting, mirrors, references, or update paths. \\

$A_{\mathrm{sw}}$ &
Signing or verification compromise &
Compromise or bypass of software signing and integrity-verification mechanisms. \\

$A_{\mathrm{sw}}$ &
Vulnerability or configuration exploitation &
Exploitation of software vulnerabilities or insecure configurations in the software supply chain. \\

\hline

$M_{\mathrm{sw}}$ &
Consumed software behavior control &
Ability to control the behavior of software consumed by downstream tools or users. \\

$M_{\mathrm{sw}}$ &
Source or repository modification &
Ability to modify source code or repository contents. \\

$M_{\mathrm{sw}}$ &
Trusted project or maintainer access &
Ability to act through trusted project or maintainer privileges. \\

$M_{\mathrm{sw}}$ &
Developer or maintainer environment control &
Ability to control or manipulate a developer or maintainer environment. \\

$M_{\mathrm{sw}}$ &
Build or toolchain execution control &
Ability to influence software execution during build or toolchain processing. \\

$M_{\mathrm{sw}}$ &
Distribution or update path control &
Ability to manipulate software distribution, retrieval, or update paths. \\

$M_{\mathrm{sw}}$ &
Signing or verification control &
Ability to manipulate software signing or verification behavior. \\

$M_{\mathrm{sw}}$ &
Vulnerability or configuration-based control &
Ability gained through exploitation of software vulnerabilities or insecure configurations. \\

\hline

$A_{\mathrm{vlsi}}$ &
IP theft, misuse, or counterfeiting &
Unauthorized extraction, use, reproduction, or counterfeiting of VLSI design assets. \\

$A_{\mathrm{vlsi}}$ &
Design tampering or malicious modification &
Unauthorized modification of VLSI design content or functionality. \\

$A_{\mathrm{vlsi}}$ &
Hardware Trojan, backdoor, or logic bomb &
Insertion or realization of malicious hidden functionality within a hardware design. \\

$A_{\mathrm{vlsi}}$ &
Information leakage or asset exposure &
Exposure or exfiltration of sensitive VLSI design information or assets. \\

$A_{\mathrm{vlsi}}$ &
Side-channel, fault, or physical attack &
Security compromise enabled through physical effects, side channels, or injected faults. \\

$A_{\mathrm{vlsi}}$ &
HLS-induced security vulnerability &
Security weaknesses introduced or exposed during high-level synthesis. \\

$A_{\mathrm{vlsi}}$ &
IC replacement or substitution &
Replacement of a legitimate IC with a malicious or counterfeit component. \\

$A_{\mathrm{vlsi}}$ &
Access abuse or system disruption &
Unauthorized access, privilege abuse, or disruption of VLSI tools or design-flow operation. \\

\hline

$S$ & Design stages &
Design specification; high-level synthesis; logic synthesis; post-synthesis verification;
design for testability; placement and routing; static timing analysis;
physical verification and signoff; post-silicon validation; manufacturing testing. \\

\hline

$O$ & Security outcomes &
Confidentiality, integrity, and availability. \\

\hline
\end{tabular}
\end{table*}

\begin{table*}[htbp]
\centering
\footnotesize
\caption{Relationships used to construct VLSI-SWSC attack paths. The complete relationship mappings used by the Prolog model are available in the accompanying Open Science artifact.}
\label{tab:model_relationships}
\begin{tabular}{p{0.20\textwidth} p{0.24\textwidth} p{0.48\textwidth}}
\hline
\textbf{Relationship} & \textbf{Model Predicate} & \textbf{Meaning} \\
\hline
\hline

$A_{\mathrm{sw}}\rightarrow M_{\mathrm{sw}}$ &
\texttt{valid\_asw\_msw\_pair} &
Maps a software supply-chain attack category to the software-side attacker capability it enables. \\

$M_{\mathrm{sw}}\rightarrow A_{\mathrm{vlsi}}$ &
\texttt{valid\_msw\_avlsi\_pair} &
Maps a software-side capability to the VLSI-side attack classes that it can realize. \\

$A_{\mathrm{vlsi}}\rightarrow S$ &
\texttt{possible\_at} &
Maps each VLSI-side attack class to the design stages where it can be introduced, enabled, observed, or propagated. \\

$A_{\mathrm{vlsi}}\rightarrow O$ &
\texttt{affects\_outcome} &
Maps each VLSI-side attack class to the confidentiality, integrity, or availability outcomes it can affect. \\

\hline
\end{tabular}
\end{table*}

\begin{table*}[htbp]
\centering
\scriptsize
\caption{Evidence-to-category mappings used in the feasibility evaluation.
The complete evidence records and mapping rationales are available in the accompanying Open Science artifact.}
\label{tab:evidence_mapping}
\begin{tabular}{c c c p{0.48\textwidth} c}
\hline
\textbf{ID} & \textbf{Evidence} & \textbf{Layer} & \textbf{Mapped Category} & \textbf{Strength} \\
\hline
\hline

M1  & E2  & $A_{\mathrm{sw}}$   & vulnerability\_or\_configuration\_exploitation & Direct \\
M2  & E2  & $M_{\mathrm{sw}}$   & consumed\_software\_behavior\_control & Direct \\
M3  & E2  & $A_{\mathrm{vlsi}}$ & design\_tampering\_or\_malicious\_modification & Indirect \\
M4  & E2  & $A_{\mathrm{vlsi}}$ & hardware\_trojan\_backdoor\_or\_logic\_bomb & Indirect \\

M5  & E3  & $A_{\mathrm{sw}}$   & vulnerability\_or\_configuration\_exploitation & Direct \\
M6  & E3  & $M_{\mathrm{sw}}$   & vulnerability\_or\_configuration\_based\_control & Direct \\

M7  & E8  & $M_{\mathrm{sw}}$   & consumed\_software\_behavior\_control & Direct \\
M8  & E9  & $M_{\mathrm{sw}}$   & consumed\_software\_behavior\_control & Direct \\

M9  & E10 & $M_{\mathrm{sw}}$   & consumed\_software\_behavior\_control & Direct \\
M10 & E10 & $A_{\mathrm{vlsi}}$ & design\_tampering\_or\_malicious\_modification & Indirect \\
M11 & E10 & $A_{\mathrm{vlsi}}$ & access\_abuse\_or\_system\_disruption & Indirect \\

M12 & E16 & $M_{\mathrm{sw}}$   & consumed\_software\_behavior\_control & Direct \\

M13 & E19 & $M_{\mathrm{sw}}$   & consumed\_software\_behavior\_control & Direct \\
M14 & E19 & $A_{\mathrm{vlsi}}$ & design\_tampering\_or\_malicious\_modification & Indirect \\

M15 & E22 & $A_{\mathrm{sw}}$   & vulnerability\_or\_configuration\_exploitation & Direct \\
M16 & E22 & $M_{\mathrm{sw}}$   & vulnerability\_or\_configuration\_based\_control & Direct \\

M17 & E27 & $M_{\mathrm{sw}}$   & consumed\_software\_behavior\_control & Direct \\
M18 & E26 & $A_{\mathrm{sw}}$   & vulnerability\_or\_configuration\_exploitation & Indirect \\

M19 & E35 & $M_{\mathrm{sw}}$   & consumed\_software\_behavior\_control & Direct \\
M20 & E34 & $A_{\mathrm{sw}}$   & vulnerability\_or\_configuration\_exploitation & Indirect \\
M21 & E28 & $M_{\mathrm{sw}}$   & consumed\_software\_behavior\_control & Indirect \\

M22 & E36 & $A_{\mathrm{sw}}$   & build\_or\_toolchain\_compromise & Direct \\
M23 & E36 & $M_{\mathrm{sw}}$   & build\_or\_toolchain\_execution\_control & Direct \\
M24 & E37 & $M_{\mathrm{sw}}$   & build\_or\_toolchain\_execution\_control & Direct \\

M25 & E48 & $A_{\mathrm{sw}}$   & build\_or\_toolchain\_compromise & Direct \\
M26 & E48 & $M_{\mathrm{sw}}$   & build\_or\_toolchain\_execution\_control & Direct \\

M27 & E38 & $M_{\mathrm{sw}}$   & consumed\_software\_behavior\_control & Direct \\

M28 & E41 & $A_{\mathrm{sw}}$   & vulnerability\_or\_configuration\_exploitation & Direct \\
M29 & E41 & $M_{\mathrm{sw}}$   & vulnerability\_or\_configuration\_based\_control & Direct \\

M30 & E44 & $M_{\mathrm{sw}}$   & consumed\_software\_behavior\_control & Direct \\
M31 & E43 & $A_{\mathrm{sw}}$   & vulnerability\_or\_configuration\_exploitation & Indirect \\
M32 & E39 & $M_{\mathrm{sw}}$   & consumed\_software\_behavior\_control & Indirect \\

M33 & E45 & $A_{\mathrm{sw}}$   & build\_or\_toolchain\_compromise & Direct \\
M34 & E45 & $M_{\mathrm{sw}}$   & build\_or\_toolchain\_execution\_control & Direct \\

M35 & E46 & $A_{\mathrm{sw}}$   & build\_or\_toolchain\_compromise & Direct \\
M36 & E46 & $M_{\mathrm{sw}}$   & build\_or\_toolchain\_execution\_control & Direct \\

M37 & E47 & $A_{\mathrm{sw}}$   & build\_or\_toolchain\_compromise & Direct \\
M38 & E47 & $M_{\mathrm{sw}}$   & build\_or\_toolchain\_execution\_control & Direct \\

M39 & E49 & $A_{\mathrm{vlsi}}$ & access\_abuse\_or\_system\_disruption & Indirect \\
M40 & E50 & $A_{\mathrm{vlsi}}$ & access\_abuse\_or\_system\_disruption & Indirect \\

\hline
\end{tabular}
\end{table*}

\begin{table*}[htbp]
\centering
\footnotesize
\caption{Evidence-based feasibility rubric for synthesized VLSI-SWSC attack paths. 
The corresponding Prolog rules implement these High, Medium, and Low feasibility conditions.}
\label{tab:feasibility_rubric}
\begin{tabular}{c c c c c}
\hline
\textbf{$A_{\mathrm{sw}}$ Evidence} &
\textbf{$M_{\mathrm{sw}}$ Evidence} &
\textbf{$A_{\mathrm{vlsi}}$ Evidence} &
\textbf{Stage Evidence} &
\textbf{Final Label} \\
\hline
\hline

\ding{51} & \ding{51} & \ding{51} & \ding{51} & High \\

\ding{51} & \ding{51} & \ding{55} & \ding{51}/\ding{55} & Medium \\

\ding{51} & \ding{55} & \ding{51}/\ding{55} & \ding{51} & Medium \\

\ding{55} & \ding{51} & \ding{51}/\ding{55} & \ding{51} & Medium \\

\ding{51}/\ding{55} &
\ding{51}/\ding{55} &
\ding{51} &
\ding{51}\footnotemark &
Medium \\

\ding{55} & \ding{55} &
\ding{51}/\ding{55} &
\ding{51}/\ding{55} &
Low \\

\hline
\end{tabular}
\footnotetext{For this Medium case, at least one of $A_{\mathrm{sw}}$ or $M_{\mathrm{sw}}$ must have evidence.}
\end{table*}

\begin{table*}[htbp]
\centering
\scriptsize
\caption{High-feasibility VLSI-SWSC attack paths. Each row reports the
Integrity (I) and Availability (A) path for a common
$A_{\mathrm{sw}}\!\rightarrow M_{\mathrm{sw}}\!\rightarrow
A_{\mathrm{vlsi}}\!\rightarrow S$ combination. The complete set of
717 synthesized paths and classifications is available in the accompanying
Open Science artifact.}
\label{tab:high_feasibility_paths}
\begin{tabular}{p{0.49\textwidth} p{0.25\textwidth} c c}
\hline
\textbf{$A_{\mathrm{sw}}\rightarrow M_{\mathrm{sw}}\rightarrow A_{\mathrm{vlsi}}$}
& \textbf{Stage $S$} & \textbf{I} & \textbf{A} \\
\hline
\hline

build/toolchain compromise $\rightarrow$ build/toolchain execution control
$\rightarrow$ design tampering/malicious modification
& High-Level Synthesis & P333 & P334 \\

& Logic Synthesis & P335 & P336 \\
& Post-Synthesis Verification & P337 & P338 \\
& Design for Testability & P339 & P340 \\
& Placement and Routing & P341 & P342 \\
& Static Timing Analysis & P343 & P344 \\
& Physical Verification and Signoff & P345 & P346 \\

\hline

build/toolchain compromise $\rightarrow$ build/toolchain execution control
$\rightarrow$ access abuse/system disruption
& High-Level Synthesis & P419 & P420 \\

& Logic Synthesis & P421 & P422 \\
& Post-Synthesis Verification & P423 & P424 \\
& Design for Testability & P425 & P426 \\
& Placement and Routing & P427 & P428 \\
& Static Timing Analysis & P429 & P430 \\
& Physical Verification and Signoff & P431 & P432 \\
& Manufacturing Testing & P435 & P436 \\

\hline

vulnerability/configuration exploitation $\rightarrow$
vulnerability/configuration-based control $\rightarrow$
design tampering/malicious modification
& High-Level Synthesis & P614 & P615 \\

& Logic Synthesis & P616 & P617 \\
& Post-Synthesis Verification & P618 & P619 \\
& Design for Testability & P620 & P621 \\
& Placement and Routing & P622 & P623 \\
& Static Timing Analysis & P624 & P625 \\
& Physical Verification and Signoff & P626 & P627 \\

\hline

vulnerability/configuration exploitation $\rightarrow$
vulnerability/configuration-based control $\rightarrow$
access abuse/system disruption
& High-Level Synthesis & P700 & P701 \\

& Logic Synthesis & P702 & P703 \\
& Post-Synthesis Verification & P704 & P705 \\
& Design for Testability & P706 & P707 \\
& Placement and Routing & P708 & P709 \\
& Static Timing Analysis & P710 & P711 \\
& Physical Verification and Signoff & P712 & P713 \\
& Manufacturing Testing & P716 & P717 \\

\hline
\end{tabular}
\end{table*}

\end{document}